\documentclass[reprint,
superscriptaddress,
amsmath,amssymb,
aps,
pra,
floatfix,
]{revtex4-2}

\usepackage{natbib}                                         
\usepackage[T1]{fontenc}                                       
\usepackage[colorlinks=true,citecolor=royalBlue,linkcolor=bordeauxRed,urlcolor=oliveGreen]{hyperref}
\usepackage{url}                                            
\usepackage{amsfonts}                                       
\usepackage{amsmath,amsthm,amssymb}                         
\usepackage{bm}                                             
\usepackage{cleveref}                                       
\usepackage{xcolor}                                         
\usepackage{physics}
\usepackage{dsfont}
\usepackage{multirow}
\usepackage{array}[=2016-10-06]
\usepackage{pifont}
\usepackage[export]{adjustbox}
\usepackage[caption=false]{subfig}
\usepackage{siunitx}
\let\svqty\qty
\usepackage{physics}
\let\qty\svqty

\definecolor{bordeauxRed}{RGB}{178, 34, 52} 
\definecolor{royalBlue}{RGB}{65,105,225} 
\definecolor{oliveGreen}{RGB}{128,128,0} 

\newcommand*\diff{\mathop{}\!\mathrm{d}}                    
\begin{document}

\title{Emergent quantum field theories on curved spacetimes in spinor Bose-Einstein condensates: from scalar to Proca fields}

\author{Christian F.\@ Schmidt}
    \thanks{Both authors contributed equally to this work. \newline Electronic addresses: \newline \href{mailto:christian.schmidt@uni-jena.de}{christian.schmidt@uni-jena.de}, \newline \href{mailto:simon.brunner@uibk.ac.at}{simon.brunner@uibk.ac.at} \newline }
    \affiliation{Theoretisch-Physikalisches Institut, Friedrich-Schiller-Universit\"{a}t Jena,\\
    Max-Wien-Platz 1, 07743 Jena, Germany}

\author{Simon Brunner}
    \thanks{Both authors contributed equally to this work. \newline Electronic addresses: \newline \href{mailto:christian.schmidt@uni-jena.de}{christian.schmidt@uni-jena.de}, \newline \href{mailto:simon.brunner@uibk.ac.at}{simon.brunner@uibk.ac.at} \newline }
    \affiliation{Institut f\"{u}r Theoretische Physik, Universit\"{a}t Heidelberg, \\ Philosophenweg 16, 69120 Heidelberg, Germany}
    \affiliation{Institut f\"{u}r Theoretische Physik, Universit\"{a}t Innsbruck, \\ A-6020 Innsbruck, Austria}

\author{Stefan Floerchinger}
    \email{stefan.floerchinger@uni-jena.de}
    \affiliation{Institut f\"{u}r Theoretische Physik, Universit\"{a}t Heidelberg, \\ Philosophenweg 16, 69120 Heidelberg, Germany}
    \affiliation{Theoretisch-Physikalisches Institut, Friedrich-Schiller-Universit\"{a}t Jena,\\
    Max-Wien-Platz 1, 07743 Jena, Germany}

\date{\today}

\begin{abstract}
    We consider excitations of a spin-1 Bose-Einstein-condensate (BEC) in the vicinity of different mean-field configurations 
    and derive mappings to emergent relativistic quantum field theories minimally coupled to curved acoustic spacetimes. 
    The quantum fields are typically identified with Nambu-Goldstone bosons, such that the structure of the analogue quantum field theories on curved spacetimes depends on the (spontaneous) symmetry breaking pattern of the respective ground-state.
    The emergent spacetime geometries are independent of each other and exhibit bi-metricity in the polar and antiferromagnetic phase, whereas one has tri-metricity in the ferromagnetic phase. 
    Compared to scalar BECs, the spinor degrees of freedom allow to investigate massive vector and scalar fields where the former is a spin-nematic rotation mode in the polar phase which can be cast into a Proca field that is minimally coupled to a curved spacetime that emerges on length scales larger than the spin-healing length.
    Finally, we specify the Zeeman couplings and the condensate trap to be spacetime-dependent such that a cosmological FLRW-metric can be achieved. 
    This work enables a pathway towards quantum-simulating cosmological particle production of Proca quanta via quenching the quadratic Zeeman-coefficient or via magnetic field ramps, which both result in the creation of spin-nematic squeezed states. 
\end{abstract}

\maketitle

\section{Introduction}
\label{sec:Introduction}
Quantum field theory in curved spacetime (QFTCS) \cite{birrell_davies_1982,Fulling1989,Wald1995,mukhanov_winitzki_2007} is a salient area of research that deals with the interaction of quantized matter with gravitational fields, 
resulting in phenomena such as cosmological particle production \cite{Parker1969} and Hawking radiation \cite{Hawking1975}.
In recent years, analogue simulators that mimic these effects of QFTCS have seen significant interest and development following the original work of Unruh \cite{Unruh1981, Unruh1995}.
Typically, the simulation platforms are optical or condensed matter systems \cite{Philbin2008,Horstmann2010,Weinfurtner2011,Euve2016,Torres2017,Drori2019,Wittemer2019,Steinhauer2022,Jacquet2022,HallerMeng2023,Tajik2023,Tajik2023AreaLaw,Jacquet2023,Shi2023,Tolosa2024,Falque2024}
in which a relativistic quantum field theory in curved spacetime emerges as an effective description for linearized field perturbations around a background that in turn shapes the curved geometry \cite{Visser1998,Barcelo2011,Visser2002,UnruhSchuetzhold2007}.
This interdisciplinary pursuit of analogue systems potentially sparks new questions and concepts for the model system \cite{Volovik2003}.

Owing to technological advances, quantum simulation of QFTCS with ultracold atoms has seen rapid development from Gedankenexperiments \cite{BarceloBEC2001,Novello2002,Barcelo2003Hawking,Fedichev2003,Visser2005,Uhlmann2005,CalzettaHu2005,Prain2010,Bilic2013} 
to modern table-top experiments \cite{Hung2013,Chen2021,Viermann2022,Sparn2024,Eckel2018,Jaskula2012,MunozDeNova2019,Hu2019,Ribeiro2022,Jenkins2024}, that exhibit precise control, high tunability and accurate readout. 
A prime example is the successive development of analogue FLRW-cosmology simulators \cite{BarceloFRW2003, BarceloBEC2003, Fedichev2004, Fischer2004, Jain2007, Weinfurtner2007BoseNova, Weinfurtner2009, Chatrchyan2021, Tolosa2022, Sanchez2022, Schmidt2024} based on the technology of an optical or magnetic Feshbach resonance \cite{Chin2010,Fedichev1996,Theis2004,Etrych2023}, 
which allows to imprint a temporal profile on the inter-atomic coupling to ultimately realize cosmological particle production experimentally in (single-component) Bose-Einstein condensates (BECs) \cite{Hung2013,Chen2021,Viermann2022,Sparn2024}. 

Compared to the scalar case, spinor BECs (extensively reviewed in \cite{Kawaguchi2012,Stamper-Kurn2013}) provide additional degrees of freedom that interact via spin-exchange collisions and are sensitive to an external magnetic field via the first and second-order Zeeman-effect.
In particular in the analogue gravity context, their multiple degrees of freedom overcome conditions that lead to a universal emergence of a uni-metric scalar QFTCS \cite{Barcelo2001, Barcelo2002} similar to two-component mixtures \cite{FischerSchuetzhold2004,Liberati2006,Liberati2006b,Weinfurtner2006,Calzetta2008,Zache2017,Giacomelli2020,Zenesini2024,Wang2025}.  

Structure formation experiments with spinor BECs can be conducted via quenches, resulting in correlations that can be probed and analyzed in terms of spin-density and nematicity, i.e.\  the vector and quadrupolar magnetic moments \cite{Carusotto2004,Sau2010}. 
Prominent phenomena such as phase transitions \cite{Sadler2006,Bookjans2011,Jacob2012}, universality \cite{Pruefer2018,Schmied2019,Lannig2023,Siovitz2025}, thermalization \cite{Pruefer2022}, formation of spin-domains \cite{Stenger1998} and spin-textures \cite{Sadler2006}, spin-squeezing \cite{Gross2011,Hamley2012,Kunkel2018,Kunkel2019,Kunkel2022}, spin-nematic order \cite{Zibold2016} and quantum entanglement \cite{Gross2011,Kunkel2018,Kunkel2022} were observed in that context.
This abundance of processes relevant for fundamental physics and cosmology combined with the possibility to extract a rich collection of observables \cite{Gross2011,Pruefer2020,Kunkel2018,Kunkel2019,Kunkel2022} render spinor BECs formidable quantum simulation platforms.
They have been proposed to simulate Bianchi-type spacetimes \cite{Calzetta2008}, false vacuum decay \cite{Billiam2022}, and recently, as simulators for bi-metric cosmological spacetimes \cite{Wang2025}.
Furthermore, an analogy between spin-nematic excitations of the spinor BEC and linearized gravity has been suggested in reference \cite{Chojnacki2024} (based on earlier work \cite{Smerald2013}).

We consider fluctuations around various mean-field ground states of a $f=1$ spinor BEC \cite{Ho1998,Ohmi1998} and show that certain low-energy fluctuations obey effective field theories that assume the form of relativistic quantum field theories of real scalar and vector field theories, which are each minimally coupled to curved, acoustic geometries as it is summarized in \cref{tab:Summary}.
The specific details that lead to this table are worked out in the main part of this manuscript.
Our findings are consistent with Bogoliubov theory \cite{Ohmi1998,Ho2000,Murata2007,Uchino2010,Kawaguchi2012}, hydrodynamical theory \cite{Lamacraft2008,Kudo2010,Yukawa2012} and the non-linear $\sigma$-models found in \cite{Zhou2001}.
Moreover, we shall in particular connect to the linearized gravity proposal \cite{Chojnacki2024}. Furthermore, our work can be considered as an extension of investigations concerning a scalar BEC \cite{Tolosa2022,Sanchez2022,Viermann2022,Schmidt2024,Sparn2024}.

\emph{The remainder of this paper is organized as follows:}
In \cref{sec:EffTheoryGeneral} a suitable effective action for the spin-1 BEC is expanded up to second order in terms of density, spin-density and spin-nematic fluctuations.
The \cref{sec:EmergentSpont,sec:EmergentExpl} constitute the main part of this work with derivations of low-energy effective field theories for excitations around the polar, ferromagnetic and anti-ferromagnetic case. 
\Cref{sec:EmergentSpont} considers the SU$(2)$-symmetric phases in absence of a magnetic field, whereas a finite magnetic field is considered in \cref{sec:EmergentExpl}.
In \cref{sec:ExpFeasible} we discuss experimental feasibility with an emphasis on the emergent Proca theory and particularize to emergent Friedmann-Lemaître-Robertson-Walker (FLRW) spacetimes that are specified via suitable constraints on the linear and quadratic Zeeman coefficients.
\Cref{sec:conclusion} concludes the paper with a summary and an outlook on future research directions.
\paragraph*{Notation.} 
Emergent spacetime indices are denoted in greek letters while spatial indices are denoted in roman letters. 
Furthermore, we choose the spacetime metric signature $(-,+,+,+)$.
Lie-Algebra indices are denoted by capitalized latin letters in the symmetric tensor representation and lowercase fracture letters in the spinor representation.
We use Einstein's sum convention where repeated indices are summed over. 
Vectors are written in bold letters and we denote the scalar product in $\mathbb{R}^3$ by $\langle \cdot, \cdot \rangle$, the cross-product by $\wedge$ and the tensor product by $\otimes$.
Furthermore, we abbreviate $\int_{t,\bm{r}} \equiv \int \mathrm{d} t \, \mathrm{d}^2 \bm{r}$ as well as where we abbreviate $\int_{\omega,\bm{k}} \equiv \int \mathrm{d}\omega/(2\pi) \diff^2 k/(2\pi)^2$ 

\section{Effective theory for fluctuations}
\label{sec:EffTheoryGeneral}

\subsection{Effective Action}
We consider a weakly interacting Bose gas with hyperfine spin $f=1$ and neglect magnetic dipole-dipole interactions which is appropriate for alkali atoms like $^{87}\mathrm{Rb}, \, ^{23}\mathrm{Na}$ or $^{7}\mathrm{Li}$ \cite{Stamper-Kurn2013}.
We shall be interested in Bose-Einstein condensed states, formed by macroscopic occupation of modes. 
Furthermore, we treat the BEC in two spatial dimensions although our calculations allow for lower dimensionality through suitably chosen trapping potentials.  
In particular, we shall also consider an external magnetic field $B(t, \mathbf{r})$ oriented parallel to the $z$-axis, and investigate its influence onto the two-dimensional orthogonal subspace. 

It is convenient for us to work with (the expectation value of) the bosonic field in the symmetric tensor representation of the SU$(2)$ hyperfine spin group (with $x=(t,\mathbf{r})$),
\begin{equation}
    \Phi(x) = \left( \begin{matrix} \Phi_+(x) & \Phi_0(x)/\sqrt{2} \\ \Phi_0(x)/\sqrt{2} & \Phi_-(x) \end{matrix} \right).
    \label{eq:FieldExpectationValue}
\end{equation}
Here the components are identical to the often employed representation $\tilde \Phi(x) = (\Phi_+(x), \Phi_0(x),\Phi_-(x))$.
Our ansatz for the effective action then reads
\begin{equation} 
    \Gamma[\Phi] = \int_{t,\bm{r}}  \{ \tr \big[\Phi^* ( \text{i} \hbar \partial_t + \tfrac{\hbar^2}{2 M} \boldsymbol{\nabla}^2 - V_\mathrm{ext} ) \Phi \big] - U(\Phi) \},      
    \label{eq:FullAction}
\end{equation}
with the effective potential
\begin{equation}
    U(\Phi) =  \left(- \mu  + \tfrac{2}{3} q \right) n + \tfrac{1}{2} c_0 n^2 + \tfrac{1}{2} c_1  \bm{F}^2  - p F_z + \tfrac{1}{2} q \mathcal{Q}_{zz},
    \label{eq:effectivePotential}
\end{equation}
where we introduced the scalar density 
\begin{equation}
    n = \text{tr}( \Phi^* \Phi ),
    \label{eq:density}
\end{equation}
the vector spin-density (or dipole moment)
\begin{equation}
    F_i =  \text{tr} ( \Phi^* \sigma_i \Phi),
    \label{eq:spindensity}
\end{equation}
as well as the symmetric and traceless spin-nematic (or quadrupolar) tensor 
\begin{equation}
\mathcal{Q}_{ij} =  \tr ( \Phi^* \sigma_i \Phi \sigma_j^T ) - \tfrac{1}{3} n \delta_{ij},
\label{eq:nematictensor}
\end{equation}
which has five linearly-independent degrees of freedom \cite{Smerald2013}. 
Here, the index $i \in \lbrace{x,y,z\rbrace}$ represents the three spatial directions and $\sigma_i$ are the corresponding Pauli matrices.
The nematic tensor \eqref{eq:nematictensor} is a useful tool to describe order and correlations in spinor BECs.
In particular the spin-nematically ordered phase has been observed experimentally \cite{Zibold2016} and its topolocical defects like $\pi$-disclinations (which are eponymous for the nematic phase \cite{deGennes1993}) and half-quantum vortices have been studied theoretically \cite{Leonhardt2000,Zhou2001} and experimentally \cite{Seo2015}.
Furthermore, its excitations can be categorized in rotation and deformation modes that admit a hydrodynamical formulation \cite{Yukawa2012}.
The quantum evolution of these modes can be analyzed from a quantum-informational perspective in terms of spin-nematic squeezing \cite{Hamley2012,Kunkel2019PRL,Kunkel2022}.

The collection of the quantities defined in \cref{eq:density,eq:spindensity,eq:nematictensor} will be referred to as the (spin) multipole moments in the following.

The action \eqref{eq:FullAction} and the spin-multipoles (\cref{eq:density,eq:spindensity,eq:nematictensor}) are invariant under U$(1)$-transformations, $\Phi \to \mathrm{e}^{\mathrm{i}\chi} \, \Phi, \, \Phi^* \to \mathrm{e}^{-\mathrm{i}\chi} \, \Phi^*$ related to particle number conservation.
Moreover, s-wave contact interactions and spin-exchange-collisions, parametrized via the coupling constants $c_0$ and $c_1$ (that depend on the scattering lengths of spin-zero and spin-two atoms \cite{Kawaguchi2012,Stamper-Kurn2013}), respectively, 
are invariant under spin rotations which act as SU$(2)$ transformations on the spinor $\Phi_{AB} \to U_A^{\hspace{0.15cm}C} U_B^{\hspace{0.15cm}D} \Phi_{CD} $.
However, the Zeeman-energy-shifts, parametrized in terms of $p$ and $q$, explicitly break this spin rotation symmetry down to $\mathrm{SO}(2)$ (if $p \neq 0$) or down to $\mathrm{SO}(2) \rtimes \mathbb{Z}_2$ (if $q \neq 0$), respectively \cite{Kawaguchi2012}.
Hence, the action \eqref{eq:FullAction} is only SU$(2)$-invariant in abscence of a magnetic field ($p=q=0$). 
In particular, under general SU$(2)$ spin-rotations, one finds that $\bm{F}$ transforms as a vector, 
\begin{equation}
    F_i \to \mathcal{R}_{ij} F_j,
\end{equation}
and $\mathcal{Q}$ as a tensor,
\begin{equation}
    \mathcal{Q}_{ij} \to \mathcal{R}_{ik} \mathcal{R}_{jl} \mathcal{Q}_{kl},
\end{equation}
in the defining representation of SO$(3)$ (sometimes called cartesian representation \cite{Yukawa2012}) where
\begin{equation}
    \mathcal{R}_{ij} = \tfrac{1}{2} \tr(\sigma_i U \sigma_j U^\dagger)
    \label{eq:GroupHomomorphism}
\end{equation}
is the two-to-one homomorphism between SU$(2)$ and SO$(3)$ \cite{Cornwell1997} (further details are collected in Appendix \ref{sec:SU2Transformation}).  
In summary, the effective action exhibits a $\mathrm{U}(1) \times \mathrm{SU}(2)$-symmetry in abscence of an external magnetic field.
At finite values of $q$ and $p$, SU$(2)$-invariance is explicitly broken while U$(1)$-invariance is still maintained, since the spin multipole moments are U$(1)$-invariant, but SU$(2)$-covariant. 

For our purposes, it is also interesting to analyze the transformation behaviour of the spin-one system in presence of a linear and quadratic Zeeman effect under time-reversal.
Time-reversal acts on the spinor-components as $(\mathcal{T} \tilde{\Phi})_\mathfrak{m} = (-1)^\mathfrak{m} \tilde{\Phi}_{-\mathfrak{m}}^*$ \cite{Kawaguchi2011} which is equivalent to $\Phi \to - \mathcal{T} \Phi \mathcal{T}^{-1}$ in the symmetric tensor representation, as it is inherited from time-reversal behavior of the Pauli matrices, $\sigma_i \to \mathcal{T}\sigma_i \mathcal{T}^{-1}  = -\sigma_i$ \cite{SakuraiNapolitano2020}.
Here we introduced the anti-unitary operator $\mathcal{T}= (\mathrm{i} \sigma_y) K$ where $K$ realizes complex conjugation of the spinor $K \Phi K^{-1} = \Phi^*$ and is also anti-unitary, $K K^\dagger = -1$ \footnote{It is useful to note that $\mathrm{i}\sigma_y$ realizes complex conjugation of any $U \in \mathrm{SU}(2)$ in a similar way, $(\mathrm{i}\sigma_y) U (\mathrm{i}\sigma_y)^{-1} = U^*$.}.
As a direct consequence, one finds that the spin-density is inverted by time-reversal, $F_i \to - F_i$ whereas the nematic tensor is time-reversal invariant \footnote{From anti-unitarity and $[T,K] = 0$ one finds that $\Phi^* \to T \Phi^* T^{-1}$ and $\sigma_i^T \to T \sigma_i^T T^{-1} = - \sigma_T$ from which the transformation behavior of interest can be directly deduced.}.
Therefore, an external magnetic field $(p \neq 0)$ explicitly breaks time-reversal symmetry of the effective action \eqref{eq:FullAction}, as one expects.
If however only a quadratic Zeeman shift is present $(q \neq 0, p =0)$, the action is time-reversal symmetric.

The linear Zeeman coefficient $p=-g\mu_B B$ is composed of the Landé hyperfine $g$-factor, the Bohr magneton $\mu_B$ and the external magnetic field $B$; 
whereas the quadratic Zeeman coefficient $q$ has contributions from an external magnetic field ($q_\mathrm{B}$, dc Zeeman effect) and from an off-resonant microwave dressing field ($q_\mathrm{mw}$, ac Zeeman-effect), resulting in $q =q_\mathrm{B} + q_\mathrm{mw}$. 
The former is given by
\begin{equation}\label{eq:quadraticZeemann}
    q_\mathrm{B} = \frac{(g \mu_B B)^2}{\hbar \omega_\mathrm{hf}},
\end{equation}
where $\omega_\text{hf}$ denotes the hyperfine frequency splitting.
The latter has the form \cite{Gerbier2006}
\begin{equation}
    q_\mathrm{mw} = \frac{\hbar \Omega^2}{4 \delta},
    \label{eq:qmw}
\end{equation}
for a linearly polarized microwave transition; here, $\Omega$ is the Rabi-frequency and the microwave detuning $\delta = \omega_\mathrm{mw} - \omega_\mathrm{hf}$ allows to change the sign of $q$ \cite{Gerbier2006,Guzman2011}.
Circularly polarized fields are possible as well, however these lead to a different expression than in \cref{eq:qmw}.

Note that we allow for a space- and time-dependent external trap potential that we assume to be isotropic,
\begin{equation} \label{eq:bg-trap}
    V_\text{ext}(t, r) = \tfrac{1}{2} M \omega^2(t) f(r),
\end{equation}
where $\omega(t)$ acts as a (possibly time-dependent) trapping frequency and $f(r)$ parametrizes the spatial profile, where for example $f(r)=r^2$ represents a harmonic trap. 

In this work we furthermore assume that the linear and quadratic Zeeman coefficients $p$ and $q$ can be varied independently (as has been for example realized in \cite{Stenger1998}). 
The latter can be tuned in modern ultracold atom experiments by ramping the external magnetic field ($q_\mathrm{B}$, \cite{Jacob2012}), or through an off-resonant microwave dressing field that can quench the quadratic Zeeman coefficient ($q_\mathrm{mw}$, \cite{Gerbier2006,Leslie2009,Bookjans2011,Hamley2012,Zhao2014,Kunkel2019,Kunkel2022}).  

\begin{figure*}
    \subfloat{\raisebox{-0.42cm}{\includegraphics[scale=0.27,valign=t]{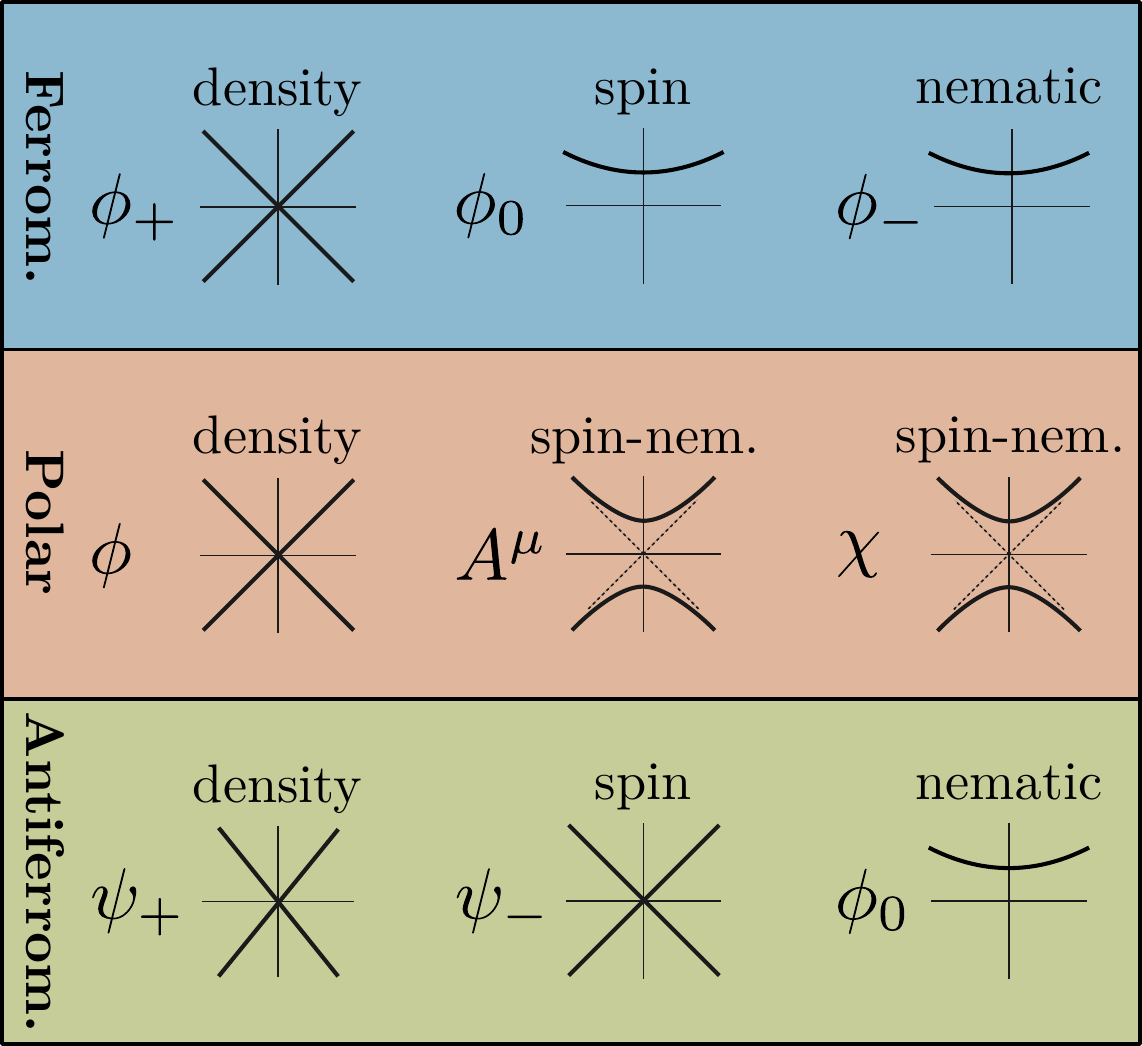}}} \hspace{0.2cm}
    \subfloat{ \includegraphics[scale=0.275,valign=t]{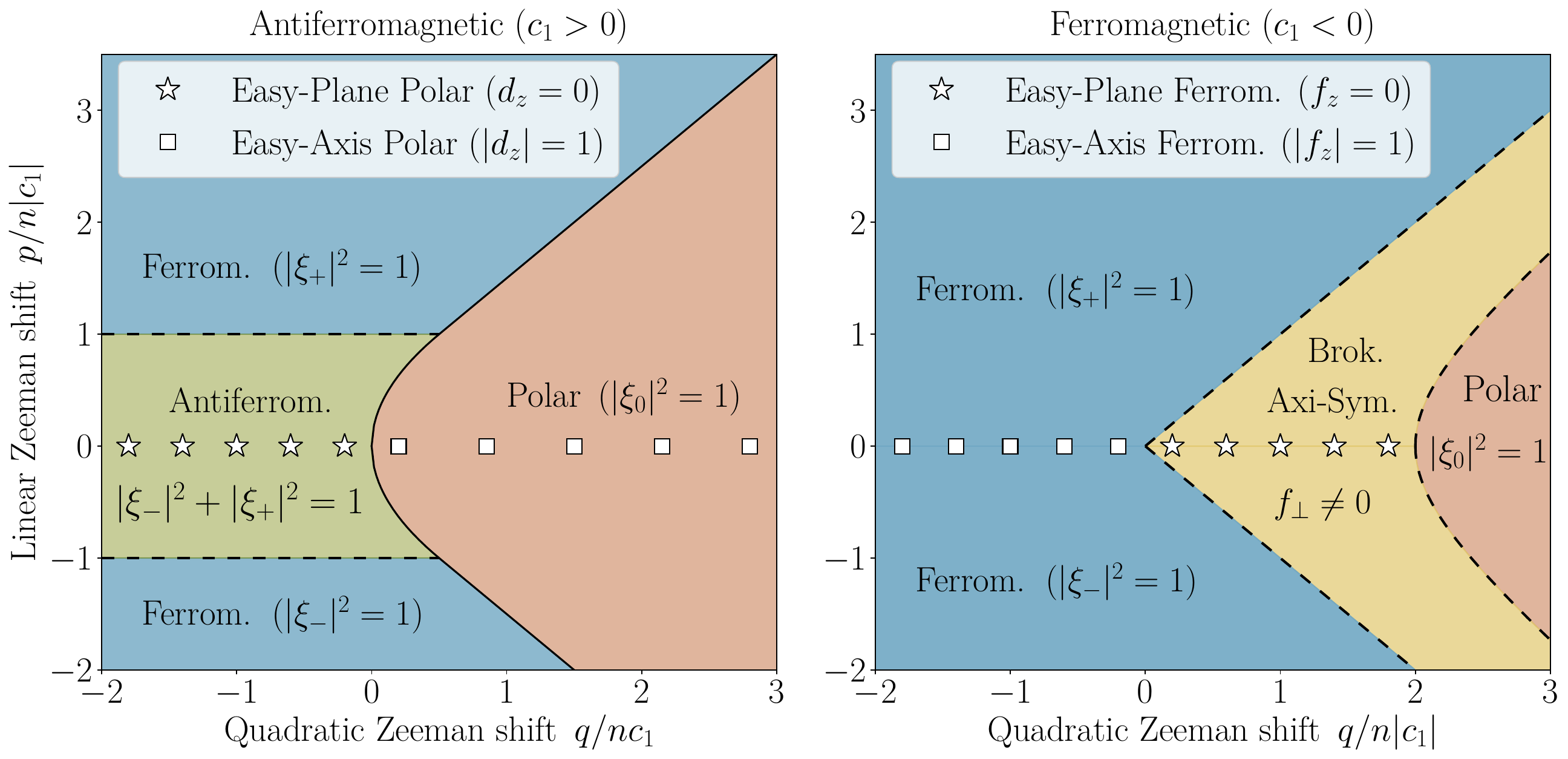}}
    
    \caption{Emergent quantum field theories in curved spacetime (left image) associated to various mean-field ground states of a spin-one BEC with antiferromagnetic (center image) and ferromagnetic (right image) interactions.
    The insets show the dispersion relations in the infrared regime, leading either to Lorentz-invariant hyperbola or Galilei-invariant parabola (in an analogue sense), depending on whether the dispersion relation is quadratic or linear in the infrared regime. 
    We notably find that transversal spin-nematic excitations of the polar state lead to an analogue vector QFTCS and a massive scalar QFTCS.
    Further details on the structure of the emergent are summarized in \cref{tab:Summary}.   
    Solid (dashed) lines in the phase-diagram indicate first (second)-order quantum phase transitions where the derivative of $U_\mathrm{mag}$ with respect to $(p,q)$ changes continously (discontinously) \cite{Kawaguchi2012}. 
    Perceptually uniform colormaps designed by \cite{Crameri2023} are used.
    \label{fig:PhaseDiagram}
    }
\end{figure*}

In the following we shall investigate effective theories around different ground-state configurations of the spinor BEC.
Finding a mean-field ground state can be understood as a minimization problem for the magnetic part of the effective potential,
\begin{equation}
   U_\mathrm{mag}(\Phi) \equiv  \frac{1}{2} c_1 \bm{F}^2  - p F_z + \frac{1}{2} q \left(Q_{zz} + \frac{4}{3}\bar n\right),
\end{equation} 
which leads to the rich quantum phase diagram of spin-1 BECs \cite{Stenger1998,Ho1998,Ohmi1998,Zhang2003,Kawaguchi2012,Stamper-Kurn2013} which is depicted in \cref{fig:PhaseDiagram}:  
Clearly, ferromagnetic interactions ($c_1 <0$) and the linear Zeeman shift $p$ favour a maximization of $\bm{F}$ (blue regions)
whereas anti-ferromagnetic interactions ($c_1 > 0$) and a positive quadratic Zeeman shift $q$ prefer a magnetically neutral state (red regions).  
The latter may still break spin-isotropy through a nematic director $\bm{d}$ that can be analyzed similar to the magnetization $\bm{F}$.
For example, both quantities may orient themselves longitudinal or transversal to the external magnetic field, which leads to easy-plane or easy-axis phases, respectively. 
The green and yellow regions can be understood as an intermediate trade-off between both agendas in the sense that a weak longitudinal alignment is present (established by a finite $p \neq 0$) at the expense of a finite transversal component.
Interestingly, the described competition may also be such that an anti-ferromagnetic BEC enters a ferromagnetic state (or vice versa). 
The former occurs if the energy scale of the positive linear ($p>0$), or negative quadratic ($q<0$) Zeeman shift is stronger than the interaction energy $\bar n c_1$. 
The broken axi-symmetric ferromagnetic phase and its excitations are not investigated in this work (consider \cite{Murata2007} for a Bogoliubov analysis).

\begin{table*}
    \renewcommand{\arraystretch}{1.25} 
    \begin{tabular}{p{1.1cm} p{1.1cm} p{4cm} p{6cm} p{2.4cm} p{2.1cm}}
        \hline 
        \hline
    Phase & Multi- \newline metric & Symmetry-Breaking\footnotemark[1] & NGB characterization\footnotemark[2]  &Physical \newline quantity& Conjugate \newline momentum \\ \hline \\[-8pt] 
    Polar & Bi-metric &
    \begin{tabular}[c]{@{}l@{}} 
    $\mathrm{U}(1) \overset{\mathrm{s}}{\longrightarrow} \mathrm{full} $ \\[2pt]
    $\mathrm{SU}(2) \overset{\mathrm{s}}{\underset{q=p=0}{\longrightarrow}} \mathrm{SO}(2) \rtimes \mathbb{Z}_2$
    \end{tabular} &
    \begin{tabular}[c]{@{}l@{}} 
    Massless scalar (linear disp.\@ $\omega_\parallel$) \\[3pt] 
    Massive vector + scalar (linear disp.\@ $\omega_\perp$) \\[4pt]
    \end{tabular} &
    \begin{tabular}[c]{@{}l@{}} 
    Phase fluc.\@  \\[3pt] 
    Nematic rot.\@ \\[4pt]
    \end{tabular} & 
    \begin{tabular}[c]{@{}l@{}} 
    Density fluc.\@  \\[3pt] 
    Spin rot.\@  \\[4pt]
    \end{tabular} \\[-5pt] \\[-5pt] \hline \\[-20pt]
    Ferrom.\@ & Tri-metric &
    \begin{tabular}[c]{@{}l@{}} 
    $\mathrm{U}(1)\overset{\mathrm{s}}{\longrightarrow} \mathrm{full}$ \\[2pt] 
    $\mathrm{SU}(2) \overset{\mathrm{s}}{\underset{q=p=0}{\longrightarrow}} \mathrm{SO}(2)$ \\[5pt]
    no association \ding{55}
    \end{tabular} &
    \begin{tabular}[c]{@{}l@{}} 
    Massless scalar (linear disp.\@ $\omega_+$) \\[3pt]
    Massive scalar (quadratic disp.\@ $\omega_0$)\footnotemark[3] \\[4pt]
    Massive scalar  (quadratic disp.\@ $\omega_-$) \\
    \end{tabular}  &
    \begin{tabular}[c]{@{}l@{}} 
        \\[-1pt]
        Phase fluc.\@  \\[3pt]
        Tran.\@ spin fluc.\@ \\[5pt]
        Nematic defor.\@ \\[-2pt] (diag.\@) \\
    \end{tabular} &
        \begin{tabular}[c]{@{}l@{}} 
        \\[-1pt]
        Density fluc.\@  \\[3pt]
        Tran.\@ spin fluc.\@  \\[5pt]
        Nematic defor.\@ \\[-2pt] (off-diag.\@) \\
    \end{tabular} 
    \\[-5pt] \\[-5pt] \hline \\[-22pt]
    \begin{tabular}{@{}l@{}} 
        Anti-\\Ferrom.\@
        \end{tabular} & Tri-metric &
    \begin{tabular}[c]{@{}l@{}} 
    \\[-2pt] $\mathrm{U}(1)\overset{\mathrm{s}}{\longrightarrow} \mathrm{full}$ \\[5pt]
    $\begin{cases}
    \mathrm{SO}(2) \rtimes \mathbb{Z}_2 \overset{\mathrm{s}}{\to} \mathbb{Z}_2 & \hspace{-0.2cm} (p = 0)  \\
    \mathrm{SO}(2) \overset{\mathrm{s}}{\to} \mathbb{Z}_2 & \hspace{-0.2cm} (p \neq 0) 
    \end{cases}$ \\[10pt]
    no association \ding{55}
    \end{tabular}  &
    \begin{tabular}[c]{@{}l@{}} 
    \\[-2pt] Massless scalar (linear dispers.\@ $\omega_\mathrm{D}$)\@\footnotemark[4] \\[12pt] Massless scalar (linear dispers.\@ $\omega_\mathrm{S}$)\@\footnotemark[4] \\[10pt]  Massive scalar (quadratic dispers.)
    \end{tabular} & 
    \begin{tabular}{@{}l@{}} 
    \\[-2pt] Phase fluc.\@  \\[12pt] Phase fluc.\@  \\[10pt] Spin-nem.\@ rot.\@
    \end{tabular} &
    \begin{tabular}{@{}l@{}} 
    \\[-2pt] Density fluc.\@  \\[12pt]  Long.\@ spin fluc.\@ \\[10pt] Spin-nem.\@ rot.\@
    \end{tabular}
    \\[10pt] \hline \hline 
    \end{tabular}
\footnotetext[1]{The s above the arrow indicates spontaneous symmetry breaking which occurs for the parameter values given below the arrow.} 
\footnotetext[2]{The mass refers to the case of explicit symmetry breaking.} 
\footnotetext[3]{Note that this QFTCS does not emerge if $p = q = 0$. For modes with quadratic dispersion relation in the infrared, the emergent theories have to be considered in the non-relativistic limit.} 
\footnotetext[4]{The exact eigenmodes are superpositions of the spin- and density modes with a dispersion relation given by the avoided crossing of $\omega_\mathrm{D}^2$ $\omega_\mathrm{S}^2$ which is explicitly shown in \cref{tab:DispRel}. }
\caption{Summary of the emergent relativistic field theories of the mean-field ground states considered in this work. This table provides further information for the visualization in \cref{fig:PhaseDiagram}. The dispersion relations of the modes are collected in \cref{tab:DispRel}. }
\label{tab:Summary}
\end{table*}

\begin{table}
    \renewcommand{\arraystretch}{1.25} 
    \begin{tabular}{p{1cm}  p{6.5cm}}
        \hline 
        \hline
    Phase & Dispersion relations \\ \hline \\[-8pt] 
    P & 
    \begin{tabular}[c]{@{}l@{}} 
    $\hbar^2 \omega_\parallel^2(k) = E_k (E_k + 2\bar n c_0)$ \\
    $\hbar^2 \omega_\perp^2(k) = E_k (E_k + 2\bar n c_1 + q)$ \\
    \end{tabular} \\[-5pt]  \\[-5pt]   \hline \\[-4pt]
    F \@ &
    \begin{tabular}[c]{@{}l@{}} 
    $\hbar^2 \omega_+^2(k) = E_k [E_k + \bar n (c_0 + c_1)]$ \\[3pt]
    $\hbar^2 \omega_0^2(k) = (E_k + p - q)^2$ \\[4pt]
    $\hbar^2 \omega_-^2(k) = (E_k + 2p - 2 \bar n c_1)^2$ \\
    \end{tabular}
    \\[-5pt] \\[-5pt] \hline \\[-22pt]
    \begin{tabular}{@{}l@{}} 
        AF \@
        \end{tabular} & 
    \begin{tabular}[c]{@{}l@{}} 
    \\[-2pt]  $\hbar^2 \omega_\pm^2(k) = \tfrac{\hbar^2 }{2}(\omega_\mathrm{D}^2(k) + \omega_\mathrm{S}^2(k))$ \\
        $\hspace{1cm} \pm \sqrt{\tfrac{\hbar^4}{4}\left(\omega_\mathrm{D}^2(k) - \omega_\mathrm{S}^2(k)\right)^2 + 4 E_k^2 f_z^2 \bar n^2 c_0 c_1}$ \\[2pt]
        where $\hbar^2 \omega_\mathrm{D,S}^2(k) = E_k (E_k + 2 \bar n c_{0,1})$ \\[8pt]
        $\hbar^2 \omega_0^2(k) = (E_k + q_- - q) (E_k + q_+ - q)$
    \end{tabular} 
    \\[10pt] \hline \hline 
    \end{tabular}
\caption{Dispersion relations of collective excitations in the considered phases.
Here we introduced atomic kinetic energy $E_k = \hbar^2 k^2 /2M$ and $q_\pm = \bar n c_1 (1 \pm \sqrt{1-f_z^2})$.
The dispersion relations are derived in \cref{sec:ExplicitSymmetryBreakingDetailed} and agree to the literature \cite{Kawaguchi2012,Yukawa2012}.}
\label{tab:DispRel}
\end{table}

\subsection{Background Field Expansion}
We split the field $\Phi(t,\mathbf{x})$ according to $\Phi = \bar \Phi + \delta \Phi$ where the background configurations of interest are stationary, radially symmetric solutions of the multi-component Gross-Pitaevskii equation
\begin{equation}
    \eval{\frac{\delta \Gamma}{\delta \Phi_{AB}^*}}_{\Phi = \bar \Phi} = 0,
    \label{eq:ExtremalCondition}
\end{equation} 
with the effective action \eqref{eq:FullAction}. Here we parametrize the ground-state manifold by 
\begin{equation}
    \begin{aligned}
        \bar \Phi (r) &= \sqrt{\bar n(r)} \, \xi = \sqrt{\bar n(r)} \left( \mqty{ \xi_+ & \xi_0 / \sqrt{2} \\ \xi_0 / \sqrt{2} & \xi_- }\right), \\
    \end{aligned}
    \label{eq:BackgroundAnsatz}
\end{equation}
where $\xi_{\pm,0}$, with $\xi_+^* \xi_+ + \xi_-^* \xi_- + \xi_0^* \xi_0 = 1$,
indicate the type of ground-state and are determined by the algebraic set of \cref{eq:Spin1GPEPlus,eq:Spin1GPE0,eq:Spin1GPEMinus}, 
which was obtained from the extremal condition \eqref{eq:ExtremalCondition} under the Thomas-Fermi approximation, where terms of the form $\hbar^2\bm{\nabla}^2 \sqrt{\bar n(r)}/2m$ are neglected by assuming sufficiently smooth density profiles.
In \cref{eq:BackgroundAnsatz} we make the single mode approximation for the background state (i.e.\@ we assume a single spatial background profile for all spin components) which is typically valid for system sizes below the spin-healing length.
However, in the Thomas-Fermi approximation (sometimes also called local density approximation) the single mode approximation gains validity on length scales above the healing length \cite{Stamper-Kurn2013} which is crucial for this work.
Therefore, we assume that the Thomas-Fermi relation, which will be derived for each mean-field ground state, holds at all times.
More details on the specific form of the multicomponent Gross-Pitaevskii equation \eqref{eq:ExtremalCondition} and its solutions can be found in Appendix \ref{sec:groundstate}. 

The fluctuations around the background configuration \eqref{eq:BackgroundAnsatz} are parametrized by six real fields,
\begin{equation} \label{eq:fluctuations}
    \delta \Phi = \frac{1}{\sqrt{2}} \left( \begin{matrix}  \phi_{+}^\text{re} + \text{i} \phi_{+}^\text{im} & \tfrac{1}{\sqrt{2}} [\phi_0^\text{re} + \text{i} \phi_0^\text{im}] \\ 
        \tfrac{1}{\sqrt{2}} [\phi_0^\text{re} + \text{i} \phi_0^\text{im}] & \phi_-^\text{re} + \text{i} \phi_-^\text{im} \end{matrix} \right).
\end{equation}

To investigate the dynamics of the fluctuations the effective action \eqref{eq:FullAction} is expanded around the background solution up to quadratic order in the fluctuating fields
\begin{equation}
    \Gamma[\Phi] = \Gamma[ \bar \Phi] + \Gamma_2[\delta \Phi ; \bar \Phi] + \mathcal{O}(\delta \Phi^3),
\end{equation}
where we already used that linear terms do not contribute due to the extremal condition \eqref{eq:ExtremalCondition}.
The effective theory for the fluctuations generally has the form 
\begin{equation}
    \begin{aligned}
        \Gamma_2 &= \int_{t,\bm{r}} \big \lbrace \tr \big [\delta \Phi^* (\mathrm{i} \hbar \partial_t + \tfrac{\hbar^2}{2m} \bm{\nabla}^2 - V_\mathrm{ext}) \delta \Phi \big] - \delta U \big \rbrace, \\
    \end{aligned}
    \label{eq:EffectiveActionFluc}
\end{equation}
where the potential energy fluctuations are 
\begin{equation}
    \begin{aligned}
        \delta U &\equiv U(\bar \Phi + \delta \Phi) - U(\bar \Phi) \\[3pt]
        &=  \left(- \mu + \bar n c_0 \right) \delta n^{(2)} + \frac{c_0}{2} \delta n^{(1)} \delta n^{(1)} + c_1 \langle \bar{\bm{F}} ,  \delta \bm{F}^{(2)} \rangle \\[3pt] 
        &+ \frac{c_1}{2}  \langle \delta \bm{F}^{(1)}, \delta \bm{F}^{(1)} \rangle - p \, \delta F_z^{(2)} + \frac{q}{2} \left(\delta \mathcal{Q}_{zz}^{(2)} + \frac{4}{3} \delta n^{(2)} \right). 
    \end{aligned}
    \label{eq:potentialFluctuations}
\end{equation} 
Here, the first- and second-order multipole fluctuations are given by
\begin{align}
        \delta n^{(1)} &\equiv \tr(\delta \Phi^* \bar \Phi) + \tr (\bar \Phi^* \delta \Phi), \label{eq:n1} \\
        \delta n^{(2)} &\equiv \tr(\delta \Phi^* \delta \Phi), \label{eq:n2} \\
        \delta F_i^{(1)} &\equiv \tr(\delta \Phi^* \sigma_i \bar \Phi) + \tr (\bar \Phi^* \sigma_i \delta \Phi), \label{eq:F1} \\
        \delta F_i^{(2)} &\equiv \tr(\delta \Phi^* \sigma_i \delta \Phi), \label{eq:F2} \\
        \delta \mathcal{Q}_{ij}^{(1)} &\equiv \tr (\delta \Phi^* \sigma_i \bar \Phi \sigma_j^T) + \tr (\bar \Phi^* \sigma_i \delta \Phi \sigma_j^T)  - \tfrac{1}{3} \delta n^{(1)}  \delta_{ij}, \label{eq:Q1} \\
        \delta \mathcal{Q}_{ij}^{(2)} &\equiv \tr(\delta \Phi^* \sigma_i \delta \Phi \sigma_j^T) - \tfrac{1}{3} \delta n^{(2)} \delta_{ij}\label{eq:Q2},
\end{align}
with explicit expressions for each phase being given in Appendix \ref{sec:physFluc}.
These multipole fluctuations are invariant under U$(1)$-transformations of the full effective spinor field, $(\bar \Phi + \delta \Phi) \to \mathrm{e}^{-\mathrm{i}\alpha} (\bar \Phi + \delta \Phi).$
Considering a SU$(2)$-transformation of the background and the fluctuations,
\begin{equation}
    \bar \Phi_{AB} + \delta \Phi_{AB} \to U_A^{\hspace{0.15cm}C} U_B^{\hspace{0.15cm}D} (\bar \Phi_{CD} + \delta \Phi_{CD}),
\end{equation}
one finds that density fluctuations (\cref{eq:n1,eq:n2}) transform as scalars, spin-density fluctuations (\cref{eq:F1,eq:F2}) as vectors and the nematic fluctuations (\cref{eq:Q1,eq:Q2}) as tensors, due to their dependence on the vector of Pauli matrices. 

Note that the second-order fluctuations and thus also the kinesial fluctuations are independent of the specific mean-field state as opposed to the potential fluctuations.
The former are given explicitly in \cref{eq:QuadraticFluctuations} while the latter, after integrating by parts, can be written as
\begin{equation}
    \begin{aligned}
    & \int_{t,\bm{r}} \tr \left[ \delta \Phi^* \left(\mathrm{i} \hbar \partial_t + \tfrac{\hbar^2}{2M} \bm{\nabla}^2\right) \delta \Phi \right]  \\ 
    &= \int_{t,\bm{r}}\left \lbrace \hbar \phi_\mathfrak{m}^\mathrm{im} \partial_t \phi_\mathfrak{m}^\mathrm{re} + \tfrac{\hbar^2}{4M} \left(\phi_\mathfrak{m}^\mathrm{re} \bm{\nabla}^2 \phi_\mathfrak{m}^\mathrm{re} + \phi_\mathfrak{m}^\mathrm{im} \bm{\nabla}^2 \phi_\mathfrak{m}^\mathrm{im}  \right) \right \rbrace
    \end{aligned}
    \label{eq:kinesialFluctuations}
\end{equation}
up to total derivatives. Note that we sum over $\mathfrak{m} \in \lbrace +1 ,0,-1 \rbrace$ in the last line. 

The results of our investigations are visualized in \cref{fig:PhaseDiagram} and summarized in \cref{tab:Summary} for the readers convenience.

\section{Emergent spacetime geometries in spontaneously broken phases}
\label{sec:EmergentSpont}

The case of a vanishing linear and quadratic Zeeman-coefficient shall be treated first, as it leads to SU$(2)$ invariance of the effective action \eqref{eq:FullAction} and ground-state configurations that spontaneously break this symmetry. 
This particularly interesting behaviour motivated pioneering experimental efforts to trap $f = 1$ atoms by purely optical means \cite{Stamper-Kurn1998,Stenger1998} which sparked multiple theoretical investigations \cite{Ho1998,Ohmi1998,Law1998,Ho2000,Zhou2001}.

\subsection{Polar (or Spin-Nematic) Phase}
\label{sec:PolarPhaseSym}
A beyond-mean-field analysis \cite{Law1998} reveals the ground-state of an anti-ferromagnetic ($c_1 > 0$) spin-1-BEC in absence of an external magnetic field to be a strongly-correlated state with super-Poissonian number fluctuations (consider \cite{Stamper-Kurn2013} for a concise summary). 
It can be considered as a fragmented condensate of spin-singlet pairs in the sense that there exist multiple macroscopic eigenvalues of the single-particle density matrix \cite{Ho2000,Mueller2006} based on the definition of Bose-Einstein condensation by Penrose and Onsager \cite{PenroseOnsager1956}.
The strict spin-isotropy and angular momentum conservation restrict the state to be a uniform superposition of mean-field ground-states over the entire Bloch sphere \cite{Mueller2006}.
However, small symmetry-violating perturbations, such as a positive quadratic Zeeman-coefficient $q > 0$, localize this state on the Bloch-sphere \cite{Barnett2010} and a mean-field description becomes applicable at low magnetic fields \cite{Ho2000,Mueller2006}.
Due to its fragility against (ambient) magnetic fields and long dynamical time scales \cite{Cui2008}, the exactly symmetric state is hard to construct experimentally \cite{Bookjans2011}.
However the vicinity of small magnetic fields can be studied via quantum quenching the quadratic Zeeman coefficient \cite{Barnett2010,Bookjans2011}. 

\subsubsection{Spin-nematic background and its fluctuations}
In the following, we will nevertheless apply mean-field theory to the background state (which captures experimentally relevant situations already at small magnetic fields \cite{Stamper-Kurn1998,Ho2000,Barnett2010}). 
For anti-ferromagnetic interactions, $c_1 >0$, the effective potential is minimized by a non-magnetized state, $\bar{\bm{F}} = 0$, such that the background is determined by 
\begin{equation}
        \left( V_\mathrm{ext} - \mu + \bar n c_0 \right) \xi_\mathfrak{m} = 0, \quad \text{for } \quad \mathfrak{m} \in \lbrace +1,0,-1 \rbrace,
\end{equation}
which only has non-trivial solutions if the Thomas-Fermi relation
\begin{equation}
    V_\mathrm{ext} - \mu + \bar n c_0 = 0
    \label{eq:ThomasFermiPolar}
\end{equation}
is fulfilled. 
Without loss of generality, any magnetically neutral background state can be parametrized by 
\begin{equation}
    \xi = \frac{\mathrm{e}^{\mathrm{i}\chi}}{\sqrt{2}} U \sigma_x U^T,
    \label{eq:RotatedPolarGS}
\end{equation} 
where $U \in \mathrm{SU}(2)$ and the global phase $\chi$ are arbitrary.
Despite the absence of magnetic order, the state at hand exhibits so-called uni-axial nematic order \cite{deGennes1993}, i.e.\@ the spin-nematic (or quadrupololar) tensor takes the form
\begin{equation}
    \bar{\mathcal{Q}} = - 2 \bar n \, (\bar{\bm{d}} \otimes \bar{\bm{d}} - \tfrac{1}{3} \mathds{1}_3).
    \label{eq:PolarNematicTensor}
\end{equation}
Note that the polar ground state is time-reversal invariant which is a necessary condition for a quantum spin-nematic \cite{Chojnacki2024}.
By analogy to ferromagnetism, the state at hand is also known as a ferroquadrupolar ground state in spin nematics \cite{Penc2010,Smerald2013,Ivanov2003,Chojnacki2024}. 
The nematic tensor \eqref{eq:PolarNematicTensor} is an orthogonal projector onto the plane perpendicular to the so-called nematic director $ \bar{\bm{d}} \equiv \mathcal{R} \bm{e}_z$ (where again the SO$(3)$ element $\mathcal{R}$ is related to the SU$(2)$ element $U$ in \cref{eq:RotatedPolarGS} through \cref{eq:GroupHomomorphism}). 
Note that the nematic tensor $\mathcal{Q}$ is invariant under the $\mathbb{Z}_2$-inversion $\bar{\bm{d}} \to -\bar{\bm{d}}$, which needs to be combined with the U(1)-transformation to realize a symmetry of the order parameter $\xi$ \cite{Kawaguchi2012}.
Therefore, the state at hand spontaneously breaks the $\mathrm{U}(1) \times \mathrm{SU(2)}$ symmetry group down to the semi-direct product $\mathrm{SO}(2) \rtimes \mathbb{Z}_2$ \footnote{Here, the semi-direct product takes into account that $\mathbb{Z}_2$ acts the elements of $\mathrm{SO}(2)$ \cite{Kawaguchi2012}.}
and the order parameter manifold is given by the coset space \cite{Zhou2001,Kawaguchi2012}
\begin{equation}
    \mathrm{U}(1) \times \mathrm{SU}(2) /(\mathrm{SO}(2) \rtimes \mathbb{Z}_2) \cong \mathrm{U}(1) \times S^2/\mathbb{Z}_2,
    \label{eq:PolarSymmetryBreaking}
\end{equation}
where S$^2$ is the three-sphere. 
It is possible to show that the spontaneous symmetry breaking pattern \eqref{eq:PolarSymmetryBreaking} leads to $1 + 2$ Nambu-Goldstone bosons (NGBs) which all have a linear dispersion relation \cite{Watanabe2012}. 
The former is the phonon (as for a scalar BEC), whereas the latter are two independent magnon modes that stem from spin-exchange collisions and induce a rotation of nematic order, as we shall show in the following. 

For the effective theory of fluctuations around the ground-state \eqref{eq:RotatedPolarGS} we find
\begin{equation}
    \begin{aligned}
        \delta n^{(2)}  V_\mathrm{ext}  +  \delta U = \frac{c_0}{2} \delta n^{(1)} \delta n^{(1)} + \frac{c_1}{2}  \langle \delta \bm{F}^{(1)} , \delta \bm{F}^{(1)} \rangle,
    \end{aligned}
\end{equation}
where we replaced the external potential according to the Thomas-Fermi relation \eqref{eq:ThomasFermiPolar}.
In the polar (or spin-nematic) phase, the density-fluctuations are scalar projections on the nematic director,
\begin{equation}
    \delta n^{(1)} = \sqrt{2 \bar n} \, \langle \bar{\bm{d}} , \Re \bm{\varphi} \rangle, 
\end{equation}
where we re-ordered the fluctuating fields in the complex vector field
\begin{equation}
\begin{aligned}
    \bm{\varphi} &\equiv \frac{1}{\sqrt{2}} (\phi_- - \phi_+) \bm{e}_x - \frac{\mathrm{i}}{\sqrt{2}} (\phi_- + \phi_+) \bm{e}_y + \phi_0 \bm{e}_z
\end{aligned}
\label{eq:FieldVector}
\end{equation}
which is of convenient use since it transforms as a cartesian vector under internal SU$(2)$-rotations and its projections onto the cartesian vectors corresponds to field-theoretic versions of the corresponding time-reversal-invariant cartesian states
(i.e.\ the eigenstates of the spin-operators $(\hat F_x, \hat F_y, \hat F_z)$ with vanishing eigenvalue 
\footnote{
    The cartesian states are defined as 
    $\ket{x} = \tfrac{1}{\sqrt{2}} (\ket{m = -1}  - \ket{m = +1}),
    \ket{y} = \tfrac{\mathrm{i}}{\sqrt{2}}  (\ket{m = -1}  + \ket{m = +1}),
    \ket{z} = \ket{0}$.
})
The spin-density fluctuations
\begin{equation}
    \delta \bm{F}^{(1)} = \sqrt{2 \bar n} \, \bar{\bm{d}} \wedge \Im \bm{\varphi} = \sqrt{2 \bar n} \, \Im \bm{\varphi}_\perp
    \label{eq:SpinDensity_NematicAFPhase}
\end{equation}
always occur in the plane perpendicular to the background nematic director $\bar{\bm{d}}$, which we parametrize with the projected field vector
\begin{equation}
        \bm{\varphi}_\perp \equiv \bm{\varphi} - \bm{\varphi}_\parallel, \qquad \bm{\varphi}_\parallel \equiv \langle \bar{\bm{d}},  \bm{\varphi} \rangle \, \bar{\bm{d}}.
    \label{eq:TransversalFieldVector_NematicAFPhase}
\end{equation}
The perturbations to the spin-nematic tensor read
\begin{equation}
    \delta \mathcal{Q}^{(1)} = \frac{\delta n^{(1)}}{\bar n} \bar{\mathcal{Q}} - \sqrt{2 \bar n} \,  \left( \bar{\bm{d}} \otimes \Re \bm{\varphi}_\perp +  \Re \bm{\varphi}_\perp \otimes \bar{\bm{d}} \right),
    \label{eq:SpinNematicFluc_P}
\end{equation}
and can be combined with the background nematic tensor in the uni-axial form
\begin{equation}
    \begin{aligned}
        \bar{\mathcal{Q}} + \delta \mathcal{Q}^{(1)} &\equiv - 2 (\bar n + \delta n^{(1)}) \\
        &\qquad \times \big[ \big(\bar{\bm{d}} + \delta \bm{d}^{(1)} \big) \otimes \big(\bar{\bm{d}} + \delta \bm{d}^{(1)} \big) - \tfrac{1}{3} \mathds{1}_3 \big],
    \end{aligned}
\end{equation}
where we identify the leading-order fluctuation of the nematic director as
\begin{equation}
    \begin{aligned}
        \delta \bm{d}^{(1)} &= \frac{\Re \bm{\varphi} \wedge \bar{\bm{d}}}{\sqrt{2 \bar n}} = \frac{\Re \bm{\varphi}_\perp}{\sqrt{2 \bar n}}, 
        \label{eq:PolarNematicTensorRotation}
    \end{aligned}
\end{equation}
indicating its rotational motion. 
As has been characterized in \cite{Yukawa2012}, the magnonic modes separate into a rotational mode of the nematic tensor (given by $\Re \bm{\varphi}_\perp$ in the present parametrization) and an amplitude oscillation of the spin-density (described by $\Im \bm{\varphi}_\perp$ here).
In particular, the presence of anisotropic, transversal spin fluctuations - despite the absence of a mean magnetization - is a characteristic feature of a spin-nematic state \cite{Ivanov2003,Laeuchli2006,Penc2010,Smerald2013,Chojnacki2024}. 
Complementary, the phonons correspond to $\bm{\varphi}_\parallel$ and are thus longitudinal to nematic-order.

In terms of the introduced fields one now finds 
\begin{equation}
        \begin{aligned}
            \Gamma_2 &= \int_{t,\bm{r}} \bigg \lbrace - \frac{\mathrm{i} \hbar}{2} \langle \bm{\varphi}^*, \partial_t \bm{\varphi} \rangle + \frac{\hbar^2}{4M} \langle \bm{\varphi}^*, \bm{\nabla}^2 \bm{\varphi} \rangle \\
            & - c_0 \bar n (\Re \bm{\varphi}_\parallel)^2 - c_1\bar n (\Im \bm{\varphi}_\perp)^2 \bigg \rbrace
        \end{aligned}
    \label{eq:PolarPhaseFluctuationSpinDOF}
\end{equation}
as an effective action that governs the dynamics of the fluctuating fields.
To derive a low-energy effective field theory for the three NGBs, we integrate out the density-fluctuation $\Re \bm{\varphi}_\parallel$ and the spin-amplitude oscillation $\Im \bm{\varphi}_\perp$ in the regime where their kinetic energy $E_k = \hbar^2 k^2/ 2M$ is neglible compared to the contact- and spin-exchange interaction energies, $E_k \ll c_0 \bar n, c_1 \bar n$.
This amounts to evaluate the action \eqref{eq:PolarPhaseFluctuationSpinDOF} on the corresponding field equations, which read
\begin{equation}
    \begin{aligned}
        - \hbar \partial_t \Im \bm{\varphi}_\parallel &= 2 \bar n c_0 \Re \bm{\varphi}_\parallel, \\
        \hbar \partial_t \Re \bm{\varphi}_\perp &= 2 \bar n c_1 \Im \bm{\varphi}_\perp,
    \end{aligned}
\label{eq:PolarSym_Eom}
\end{equation} 
in the low-energy regime. 
\subsubsection{Phonon sector: Massless, relativistic scalar field on curved spacetime}
As a result, one finds that the effective theory for the U$(1)$-NGB can be written as 
\begin{equation}
    \Gamma_2^\parallel = - \frac{\hbar^2}{2} \int_{t,\bm{r}} \sqrt{g_0} \, g_0^{\mu \nu} \partial_\mu \phi \partial_\nu \phi
    \label{eq:PhononActionPolarRelativistic}
\end{equation}
where we rescaled the field to have a relativistic mass dimension, i.e.\@ $\phi = \Im \bm{\varphi}_\parallel /\sqrt{2M}$, and introduced the acoustic metric
\begin{equation}
    (g_0)_{\mu \nu} = \left( \begin{matrix} -1 & 0 \\ 0  & c_{\mathrm{s},0}^{-2} \delta_{ij} \end{matrix} \right), \quad c^2_\mathrm{s,0}(t, r) = \frac{\bar n(r) c_0(t)}{M},
    \label{eq:AcousticMetricParallelPolar}
\end{equation}
with the abbreviation $$\sqrt{g_0} \equiv [- \det (g_0)_{\mu \nu}]^{1/2} = 1/c_\mathrm{s,0}^2.$$ 
This emergent relativistic theory is equivalent to the one for the $\mathrm{U}(1)$-NGB of a scalar BEC which was extensively studied theoretically \cite{Tolosa2022,Schmidt2024} and experimentally \cite{Viermann2022,Sparn2024}.

\subsubsection{Magnon sector: Massless, relativistic vector field on curved spacetime}
For the two SU$(2)$-NGBs we find the low-energy effective field theory 
\begin{equation}
    \Gamma_2^\perp = - \frac{\hbar^2}{4M} \int_{t,\bm{r}} \left \lbrace - \frac{M}{\bar n c_1} (\partial_t \Re \bm{\varphi}_\perp)^2 + (\bm{\nabla} \Re \bm{\varphi}_\perp)^2  \right \rbrace,
    \label{eq:SpinWavesActionSpont}
\end{equation} 
which generates the field equation 
\begin{equation}
    - \partial_t^2 \Re \bm{\varphi}_\perp + 2 \frac{\partial_t c_\mathrm{s,1}}{c_\mathrm{s,1}} \partial_t \Re \bm{\varphi}_\perp + c_\mathrm{s}^2(t) \bm{\nabla}^2 \Re \bm{\varphi}_\perp = 0,
    \label{eq:MagnonFieldEquation}
\end{equation}
that agrees with the spin-wave equation discussed in \cite{Ho1998}.
Here we introduced the propagation speed for spin-waves,
\begin{equation}
    c_{\mathrm{s},1}^2(t) = \bar n c_1(t) /M
    \label{eq:SpinWaveSpeed}
\end{equation}
The effective field theories for the phonon-sector \eqref{eq:PolarPhaseFluctuationSpinDOF} and the magnon-sector \eqref{eq:SpinWavesActionSpont} have been previously derived in \cite{Zhou2001} where the latter was already described as an $O(3)$ non-linear sigma model for a quantum spin-nematic state which was expressed in covariant notation.
In presence of a magnetic field (which also will be studied in this work) the reference identified an O(2) non-linear sigma model.

To bring \cref{eq:MagnonFieldEquation} into a relativistic covariant form we concentrate on situations where $\bar n c_1$ is spatially constant. We decompose
\begin{equation}
\frac{1}{\sqrt{2M}} \Re (\varphi_\perp)_j(t, \mathbf{x}) = \frac{\partial_j}{\sqrt{-\partial_1^2-\partial_2^2}} \chi(t, \mathbf{x}) + A_j(t, \mathbf{x}),    
\label{eq:VectorFieldDecomposition}
\end{equation}
where $\partial_j A_j(t, \mathbf{x}) = 0$, and the square root of the Laplacian can be defined with appropriate boundary conditions through a Fourier expansion.
The longitudinal magnonic degree of freedom satisfies the Klein-Gordon equation 
\begin{equation}
    \partial_\mu (\sqrt{g_1} g_1^{\mu \nu} \partial_\nu \chi) = 0
\end{equation}
where the metric has the same form as in eq.\ \eqref{eq:AcousticMetricParallelPolar},
\begin{equation}
    (g_1)_{\mu \nu} = \left( \begin{matrix} -1 & 0 \\ 0 &  c_{\mathrm{s},1}^{-2} \delta_{ij} \end{matrix} \right),
    \label{eq:AcousticMetricSpinWaves}
\end{equation}
but with the spin-wave speed $c_{\mathrm{s},1}(t)$ defined in \cref{eq:SpinWaveSpeed}.
A relativistic form for the transverse magnonic degree of freedom can be obtained by introducing the augmented vector field
\begin{equation}
    (A^\mu) = \begin{pmatrix} 0 \\ A_i \end{pmatrix},
\end{equation}
such that the field equation reads
\begin{equation}
    \frac{1}{\sqrt{g_1}}\partial_\mu \left(\sqrt{g_1} g_1^{\mu \nu} \partial_\nu A^\sigma \right) = 0,
\end{equation}
where the vector field satisfies both the Weyl gauge condition $A_0 = 0$ and the Coulomb gauge condition $\boldsymbol{\nabla} \cdot \mathbf{A} = 0$. 
As a consequence of $\partial_i \sqrt{g_1} = 0$, the covariant gauge constraint
\begin{equation}
    \nabla_\mu A^\mu \equiv \partial_\mu A^\mu + \frac{\partial_\mu \sqrt{g_1}}{\sqrt{g_1}} A^\mu = 0
    \label{eq:SubsidiarityConstraint}
\end{equation}
is fulfilled. With this, the equation of motion can be written in the form
\begin{equation}
    \nabla_\mu F^{\mu \nu} = \frac{1}{\sqrt{g_1}} \partial_\mu \left(\sqrt{g_1} F^{\mu \nu} \right) = 0,
     \label{eq:ProcaTargetFieldEq}
\end{equation}
where we introduced the field strength tensor 
\begin{equation}
    F_{\mu \nu} = \partial_\mu A_\nu - \partial_\nu A_\mu.
\end{equation}
Given that the equations of motion are as for a Abelian gauge field \cite{Heisenberg2019} one might expect that also the action be expressied in the corresponding form. We plan to investigate this point further in future work.
When considering a finite magnetic field in \cref{sec:EmergentExpl}, the spin nematic rotation mode (and thus the emergent vector field theory) will gain a mass such that the subsidiarity constraint \eqref{eq:SubsidiarityConstraint} will be automatically fulfilled. 

\subsubsection{Transformation of spin-nematic fluctuations under broken symmetry}

Before we finalize this section, we would like to briefly comment on the lattice approach \cite{Chojnacki2024,Smerald2013} where the anti-ferromagnetic Heisenberg model augmented by spin-exchange interactions \cite{Ivanov2003,Laeuchli2006,Penc2010} is used to model the spin-nematic phase at hand. 
The study concluded in the emergence of a relativistic spin-two field given by the spin-nematic fluctuations which in the present work are formulated via
\begin{equation}
    \delta Q = \delta \mathcal{Q} - \delta n^{(1)} \bar{\mathcal{Q}} / \bar n
    \label{eq:NematicFlucRelative}
\end{equation}
where $\delta \mathcal{Q}$ is given in \cref{eq:SpinNematicFluc_P}.
Our calculations reveal the non-linear sigma model used there, which is plausible since the spin-1-BEC is a member of the Heisenberg universality class \cite{Sachdev2011}, such that quantum rotor models \cite{Barnett2010} or non-linear sigma models \cite{Zhou2001,Ivanov2003} are applicable. 
However, we disagree with the interpretation as a tensor field. The equations of motion of the spin-nematic fluctuations indeed take the form
\begin{equation}
    - \partial_t^2 \delta Q_{iz} + c_{\mathrm{s},1}^2 \bm{\nabla}^2 \delta Q_{iz} = 0, \qquad i = x,y, 
\end{equation}
if we choose the background nematic director to be $\bar{\bm{d}} = \bm{e}_z$ without loss of generality.
The field equations seem to agree with those of linearized gravity with propagation speed $c_\mathrm{s,1}$ and propagation direction $\bm{d}$ (if we keep $\bar n c_1$ constant in space and time). 
Moreover, the transversal, traceless nature of $\delta Q_{ij}$ is superficially analogous to tensor metric perturbations in General Relativity \cite{Carroll2019}.

However, according to \cref{eq:SpinNematicFluc_P}, the spin-nematic perturbations decompose into the tensor product of a background part $\bm{\bar{d}}$ and a two-dimensional vector fluctuation $\Re \bm{\varphi}_\perp$.
More concretely, $\delta Q$ from \cref{eq:NematicFlucRelative} has the explicit form
\begin{equation}
    \delta Q_{ij}  =  - \sqrt{8\bar n} \, \bar{d}_i \, (\mathrm{Re} \, \varphi_\perp)_j,
    \label{eq:NematicFlucRelativeExplicit}
\end{equation}
where the background nematic director $\bm{\bar{d}}$ is fixed up to $\mathbb{Z}_2$-inversion, which would correspond to a reversal of the propagation direction. 
According to the guiding principle of Wigner \cite{Wigner1931}, the particle content of the theory is classified according to the transformation behaviour of the quantum field under the little group (or stabilizer subgroup) associated to the particles four-momentum.
In an analogue system, we identify this group with the little group of the order parameter which is $\mathrm{SO(2)} \rtimes \mathbb{Z}_2$ in the polar phase.
Considering a general element $\mathcal{R} \in \mathrm{SO(2)} \rtimes \mathbb{Z}_2$, the nematic fluctuations in the transversal plane transform in the cartesian (or defining) representation of the spin group according to
\begin{equation}
    \delta Q_{iz} \to \mathcal{R}_{ij} \mathcal{R}_{zm} \delta Q_{jm}.
\end{equation}
Now, any element of $\mathrm{SO(2)} \rtimes \mathbb{Z}_2$ can decomposed into a $(n\pi)$-rotation around the x-axis (where $n=0,1$), followed by a rotation by an angle $\alpha \in [0,2\pi)$ around the z-axis \cite{Kawaguchi2012}. 
For such elements one finds $\mathcal{R}_{zm} = (-1)^n \delta_{zm}$ and that the $\mathbb{Z}_2$-induced group element $\mathcal{R}_\perp$ in the transversal plane is either a reflection along an origin-line tilted from the x-axis by an angle $\alpha/2$,
\begin{equation}
    \mathcal{R}_\perp^{(0)} = \mqty(\cos(\alpha) & \sin(\alpha) \\ \sin(\alpha) & - \cos(\alpha)) \quad (\mathrm{if} \, n=0),
\end{equation}
or a rotation by $\alpha$,
\begin{equation}
    \mathcal{R}_\perp^{(1)} = \mqty(\cos(\alpha) & -\sin(\alpha) \\ \sin(\alpha) & \cos(\alpha)) \quad (\mathrm{if} \, n=1).
\end{equation}
Consequently we have that the nematic fluctuations of the polar phase (and thus the analogue Maxwell or Proca field) transform as vectors in the cartesian representation of the little group $\mathrm{SO}(2) \rtimes \mathbb{Z}_2$ according to 
\begin{equation}
    \delta Q_{iz} \to (-1)^n (\mathcal{R}^{(n)}_\perp)_{ij} \delta Q_{jz}.
\end{equation}

In contrast, a gravitational wave with helicity two would rotate with a phase $\mathrm{e}^{2 \mathrm{i}\vartheta}$ around its direction of propagation.

\subsection{Ferromagnetic Phase}
At vanishing external fields, the ferromagnetic phase can only occur for ferromagnetic interactions ($c_1< 0$) that favor a magnetized state.
The state at hand was accessed experimentally in the quantum-degenerate regime \cite{Guzman2011}.
Without loss of generality we model the ferromagnetic background by the spin-rotated positively polarized state
\begin{equation}
    \xi = \frac{\mathrm{e}^{\mathrm{i}\chi}}{2}  U (\mathds{1}_2 + \sigma_z) U^T,
\end{equation}
where $U \in \mathrm{SU}(2)$ and the global phase $\chi$ is arbitrary. The mean magnetization is $\bar{\bm{F}} = \bar n \bar{\bm{d}}$,
where we chose $\bar{\bm{d}}$ as the unit vector pointing into $z$-direction. 
Note that the oppositely polarized case can be investigated analogously with a few reparametrizations \footnote{One has to map $\bm{F} \to - \bm{F}$ and $\phi_\pm^\mathrm{re} \to \phi_\mp^\mathrm{re}$ as well as $\phi_\pm^\mathrm{im} \to - \phi_\mp^\mathrm{im}$ and finally $\mathrm{e}_\pm \to \mathrm{e}_\mp$}. 
In contrast to the polar state, the ferromagnetic state is not time-reversal invariant as can be deduced from its finite magnetization.
The nematic tensor is again a projector on the xy-plane,
\begin{equation}
    \bar{\mathcal{Q}} = \bar n (\bar{\bm{d}} \otimes \bar{\bm{d}} - \tfrac{1}{3} \mathds{1}_3),
\end{equation}
however, the $\mathds{Z}_2$-inversion symmetry is spontaneously broken in the ferromagnetic state through the finite background magnetization, such that its little group is only $\mathrm{SO}(2)$ \cite{Kawaguchi2012}.
Hence, the spontaneous symmetry breaking is described by the coset space
\begin{equation}
    \mathrm{U}(1)  \times  \mathrm{SU}(2) / \mathrm{SO}(2) \cong \mathrm{U}(1) \times S^2,
\end{equation} 
which differs from the polar coset space \eqref{eq:PolarSymmetryBreaking} by the absence of $\mathbb{Z}_2$-equivalent configurations.
This affects the NGBs such that there is only one with linear dispersion (i.e.\@ the phonon) and another single one with quadratic dispersion \cite{Watanabe2012},
As we shall show, the latter corresponds to a rotation of the nematic tensor which is accompanied by the rotation of the magnetization \cite{Yukawa2012}.
In the following it will be convenient to introduce
\begin{equation}
    \begin{aligned}
        \bm{\varphi} &\equiv \tfrac{1}{\sqrt{2}} \left((\phi_+ - \phi_-) \bm{e}_x + \mathrm{i} (\phi_+ + \phi_-) \bm{e}_y \right), \quad \bm{\varphi}_0 \equiv \phi_0 \bm{e}_z, 
    \end{aligned}
    \label{eq:FieldVectorFerro}
    \end{equation} 
such that 
\begin{equation}
\begin{aligned}
    \delta n^{(1)} &= \sqrt{\bar n} \, \Re \,  \langle \bm \varphi - \bm \varphi_0, \bm{e}_x - \mathrm{i} \bm{e}_y  \rangle = \sqrt{2\bar n} \phi_+^\mathrm{re}.
\end{aligned}
    \label{eq:DensityFlucFerro}
\end{equation}
In contrast to the polar phase, the nematic director is projected out in the density-fluctuation \eqref{eq:DensityFlucFerro}. 
For the spin-density fluctuations, we find 
\begin{equation}
    \begin{aligned}
        \delta \bm{F}^{(1)} &= \sqrt{\bar n} \, \Im \left[(\bm{\varphi} + \bm{\varphi}_0) \wedge (\bm{e}_x - \mathrm{i} \bm{e}_y) \right] \\
        &= \sqrt{\bar n} (\phi_0^\mathrm{re} \bm{e}_x + \phi_0^\mathrm{im} \bm{e}_y + \sqrt{2} \phi_+^\mathrm{re} \bm{e}_z). \\[3pt]
    \end{aligned}
    \label{eq:SpinDensityFlucFerro}
\end{equation}  
Hence, the real and imaginary part of the perturbation $\phi_0$ lead to spin-density fluctuations in the plane-orthogonal to $\bar{\bm{F}}$, inducing a rotational motion, 
whereas $\phi_+^\mathrm{re}$ is the longitudinal spin-excitation which is known to be identical with the density-excitation of the ferromagnetic state since the system is fully magnetized \cite{Ohmi1998,Kawaguchi2012}. 
The spin-nematic fluctuations are again obtained via symmetrized tensor products of two vectors whose overlap gives the density-fluctuation \eqref{eq:DensityFlucFerro}, 
\begin{equation}
    \begin{aligned}
        \delta \mathcal{Q}^{(1)} &=  \frac{\delta n^{(1)}}{\bar n} \bar{\mathcal{Q}} + \delta n^{(1)} (\bm{e}_x \otimes \bm{e}_x + \bm{e}_y \otimes \bm{e}_y)  \\
        &+ \sqrt{\bar n} \Re \left[(\bm \varphi - \bm{\varphi}_0) \otimes \bm{e}_-  + \bm{e}_- \otimes (\bm \varphi - \bm{\varphi}_0) \right]. 
    \end{aligned}
\end{equation}
where $\bm{e}_- = \bm{e}_x - \mathrm{i} \bm{e}_y$.
Combining the fluctuation with the background, we find the manifestly traceless expression
\begin{equation}
    \begin{aligned}
    &\frac{\mathcal{Q} + \mathcal{Q}^{(1)}}{\sqrt{2 \bar n}}= \frac{\bar n + \delta n^{(1)}}{\sqrt{2 \bar n}} \left(\bm{d}_\parallel \otimes \bm{d}_\parallel - \tfrac{1}{3} \mathds{1}_3\right) \\ 
    &+ \phi_-^\mathrm{im} (\bm{e}_x \otimes \bm{e}_y + \bm{e}_y \otimes \bm{e}_x ) + \phi_-^\mathrm{re} (\bm{e}_x \otimes \bm{e}_x - \bm{e}_y \otimes \bm{e}_y ),
    \end{aligned}
    \label{eq:NematicFlucFerro}
\end{equation}
where we introduced the rotated background director 
\begin{equation}
   \bm{d}_\parallel = \bar{\bm{d}} - \frac{\Im (\phi_0 \bm{e_-})}{2 \sqrt{\bar n}} \wedge \bar{\bm{d}}.  
   \label{eq:RotatedDirectorFerro}
\end{equation}
Consequently, the perturbation $\phi_0$ induces a rotation of the nematic director as for the polar phase.
However, there is also a deformation mode ($\phi_-$) which leads to the less-symmetrical bi-axial shape in contrast to the polar phase (where uni-axial order is preserved by the perturbations).
For the effective potential of the fluctuations, we find
\begin{equation}
    \begin{aligned}
        &\delta U + V_\mathrm{ext} \delta n^{(2)}  \\
        &= \tfrac{1}{2} c_0 (\delta n^{(1)})^2  + \tfrac{1}{2} c_1 \big( \delta \bm{F}^{(1)} \big)^2 + \bar n c_1 (\delta \bm{F}_z^{(2)} \mp \delta n^{(2)}) \\          
        &= \bar n (c_0 + c_1) (\phi_+^\mathrm{re})^2  - \bar n c_1\left( (\phi_-^\mathrm{re})^2 + (\phi_-^\mathrm{im})^2 \right)  ,
    \end{aligned}
    \label{eq:EffectivePotentialFerroFluc}
\end{equation}
where we used that the Thomas-Fermi relation in the ferromagnetic phase reads
\begin{equation}
    \bar n (c_0 + c_1) + V_\mathrm{ext} =  \mu.
\end{equation}

\subsubsection{Phonon sector: Massless, relativistic scalar field on curved spacetime}
Integrating out the density fluctuation $\phi_+^\mathrm{re}$ in the acoustic limit $E_k \ll \bar n (c_0 + c_1)$, we find
an effective action for the $\mathrm{U}(1)$-NGB of the ferromagnetic phase,
\begin{equation}
    \begin{aligned}
        \Gamma_2^+ &= - \frac{\hbar^2}{2} \int_{t,\bm{r}} \sqrt{g_+} \, g_+^{\mu \nu} \partial_\mu \phi_+ \partial_\nu \phi_+,
        \label{eq:PhononActionFerro}
    \end{aligned}
\end{equation}
where we introduced the analogue relativistic field $\phi_+ = \phi_+^\mathrm{im}/\sqrt{2M}$ and the acoustic metric
\begin{equation}
        ({g_+})_{\mu \nu}  = \left( \begin{matrix} -1 & 0 \\ 0  & c_{\mathrm{s},+}^{-2} \delta_{ij} \end{matrix} \right),
    \label{eq:AcousticMetricFerro}
\end{equation}
with the sound-speed
\begin{equation}
    c^2_\mathrm{s,+}(t, r) = \frac{\bar n(r)}{M} \left(c_0(t) + c_1(t)\right).
\end{equation}
In contrast to the polar phase, where each type of interaction leads to a separate NGB with linear dispersion, 
here both the s-wave contact interaction energy as well as the spin interaction energy contribute to the propagation of the single NGB with linear dispersion because the background is fully magnetized \cite{Kawaguchi2012}. 

\subsubsection{Magnon sector}
The rotational modes of the spin-density are identical to the rotational modes of the nematic director, i.e.\@ $\delta F_x^{(1)} = \delta \mathcal{Q}_{xz}^{(1)}$ and $\delta F_y^{(1)} = \delta \mathcal{Q}_{yz}^{(1)}$, and read
\begin{equation}
    \delta F_x^{(1)} = \sqrt{\bar n} \, \phi_0^\mathrm{re}, \quad
    \delta F_y^{(1)} = \sqrt{\bar n} \, \phi_0^\mathrm{im},
    \label{eq:TransversalSpinDensityFlucFerro}
\end{equation}
in terms of the fluctuations of the order parameter. 
The equations of motion are those of the real and imaginary parts of a free Schrödinger field
\begin{equation}
    \hbar \partial_t \phi_0^\mathrm{re} =  \frac{\hbar^2}{2M} \bm{\nabla}^2 \phi_0^\mathrm{im},  \quad -\hbar \partial_t \phi_0^\mathrm{im} = \frac{\hbar^2}{2M} \bm{\nabla}^2 \phi_0^\mathrm{re},
    \label{eq:Phi0EoMFerro} 
\end{equation}
The presence of an external field (to be analyzed in \cref{sec:EmergentExpl}) will induce a potential fluctuation thereby providing an energy scale relative to which an effective theory can be defined.

\subsubsection{Nematic sector: Non-relativistic limit}
\label{subsubsec:NemSecNR}
Integrating out the diagonal nematic deformation mode $\phi_-^\mathrm{re}$ (cref.\@ \cref{eq:NematicFlucFerro}) in the low-energy regime $E_k \ll 2 \bar n \lvert c_1 \rvert$, 
we find
\begin{equation}
    \begin{aligned}
    \Gamma_2^- = - \frac{1}{2} \int_{t,\bm{r}} \frac{1}{2 M c_-^2} \bigg \lbrace & - (\hbar \partial_t \phi_-^\mathrm{im})^2 + c_{\mathrm{s},-}^{2} \left(\hbar\bm{\nabla} \phi_-^\mathrm{im} \right)^2 \\
    &+ (2 \bar n c_1)^2 (\phi_-^\mathrm{im})^2 \bigg \rbrace,
    \end{aligned}
    \label{eq:EA_NematicDeformFerro}
\end{equation}
where we introduced the characteristic speed 
\begin{equation}
    c_{\mathrm{s},-}^2 = \lvert n c_1 / M \rvert
\end{equation}
This can be written as a QFTCS described by the effective action 
\begin{equation}
    \begin{aligned}
        \Gamma_2^-  = - \frac{\hbar^2}{2}  \int_{t,\bm{r}} \sqrt{g_-} \, \left(g_-^{\mathrm{\mu \nu}} \partial_\mu \phi_- \partial_\nu \phi_- - m_\phi^2 \phi_-^2 \right),
        \label{eq:Spin2ActionFerro}
    \end{aligned}
\end{equation}
in terms of the relativistic field $\phi_- = \phi_-^\mathrm{im} / \sqrt{2M}$, the acoustic metric
\begin{equation}
    (g_-)_{\mu \nu} = \left( \begin{matrix} 1 & 0 \\ 0  & c_{\mathrm{s},-}^{-2} \delta_{ij} \end{matrix} \right)
\end{equation}
and the relativistic mass
\begin{equation}
    m_\phi^2 = \frac{4M^2}{\hbar^2} c_\mathrm{s,-}^4 = \frac{4}{\hbar^2} \bar n^2(r) c_1^2(t).
    \label{eq:mPhiNematicSpont}
\end{equation}
However, consistency with the appromation $E_k \ll 2 \bar n \lvert c_1 \rvert$ requires the consider the dispersion relation $\hbar^2 \omega_-^2(k) = (E_k + 2 \bar n \lvert c_1\rvert)^2$ for small momenta where it depends quadratically on $k$.
Consequently, the emergent theory is only valid in the \emph{non-relativistic} limit.
To see this explicitly, let us express $\Gamma_2^-$ in Fourier-space,
\begin{equation}
    \begin{aligned}
            \Gamma_2^- = \frac{\hbar^2}{2} \int_{\omega,\bm{k}} \frac{\omega^2 - \omega_-^2(k)}{2M c_-^2}  \phi^\mathrm{im}_-(-\omega,-\bm{k}) \phi^\mathrm{im}_-(\omega,\bm{k}),
    \end{aligned}
    \label{eq:EA_NematicDeformFerroDisp}
\end{equation}
where we identify
\begin{equation}
    \begin{aligned}
        \hbar \omega_-(k) &= \sqrt{(\hbar k c_- )^2 + (2 \bar n c_1)^2} \approx m_- c_-^2 + \frac{\hbar^2 k^2}{2m_-}
    \end{aligned}
\end{equation}
at leading order in $\hbar k / m_- c_-$ and introduced the rest-mass term $m_- c_-^2 = 2 \bar n \lvert c_1 \rvert$ which is consistent with the field-mass \eqref{eq:mPhiNematicSpont} up to a choice of units, $m_\phi = m_- c_-^2/\hbar^2$.
The relativistic limit of the real Klein-Gordon field on a FLRW-spacetime was studied in detail in \cite{Heyen2020} by generalizing a transformation to a non-relativistic Schrödinger field \cite{Namjoo2018} to curved spacetimes. 
The developed methods could also be applied in the present case, which we consider beyond the scope of this work.

\section{Emergent Spacetime Geometries in Explicitly Broken Phases}
\label{sec:EmergentExpl}
In the following we shall discuss the implications of a finite magnetic field ($p \neq 0, q \neq 0$) which conventionally shall be pointing into the $z$-direction. 
The remaining symmetry of the ground state will be the same as for the spontaneously broken phases; however, the spin rotation group SU$(2)$ is explicitly broken by the external magnetic field such that the $\mathrm{SU}(2)$-NGBs acquire a mass, whereas the U$(1)$-symmetry is still spontaneously broken, resulting in the presence of a massless phonon.
More detailed calculations can be found in Appendix \ref{sec:ExplicitSymmetryBreakingDetailed}.

\subsection{Polar (or Spin-Nematic) Phase}
\label{sec:PolarPhase}
In the presence of an external magnetic field, the polar phase can be maintained if $q>0$ for anti-ferromagnetic interactions ($c_1>0$) as positive values of both couplings naturally favour a magnetically neutral state.
However, even for ferromagnetic interactions ($c_1<0$) the polar phase can be maintained if the quadratic Zeeman-shift is large compared to the ferromagnetic interaction energy, $q > \bar n c_1$.
Then ground-state is explicitly given by $\xi = \sigma_x /\sqrt{2}$ and the background nematic director aligns with the external magnetic field, $\bar{\bm{d}} = \bm{e}_z$.
Therefore, the phase at hand is still symmetric under O$(2)$-transformations in the plane perpendicular to $\bar{\bm{d}} \parallel \bm{B}$.

The effective theory for the phonon is not affected by the external magnetic field and therefore still described by the relativistic scalar field theory \eqref{eq:PhononActionPolarRelativistic}.
The rotation mode of the nematic director is now massive due to the explicit SU$(2)$-symmetry breaking and written as the relativistic vector field 
\begin{equation}
    \begin{aligned}
        \Gamma_2^\perp &= - \frac{\hbar^2}{4M} \int_{t,\bm{r}} \bigg \lbrace - \frac{M}{\bar n c_1 + q/2} (\partial_t \Re \bm{\varphi}_\perp)^2  \\
        &+ (\bm{\nabla} \Re \bm{\varphi}_\perp)^2 +  \frac{2M}{\hbar^2} q \, (\Re \bm{\varphi}_\perp)^2  \bigg \rbrace.
    \end{aligned}
    \label{eq:NematicRotationMassive}
\end{equation}
We decompose the vector field $\Re \bm{\varphi}_\perp$ into a longitudinal and a transversal part as in \cref{eq:VectorFieldDecomposition} and obtain the massive Klein-Gordon equation
\begin{equation}
    \frac{1}{\sqrt{g_1}} \partial_\mu \left(\sqrt{g_1}g^{\mu \nu} \partial_\nu \chi \right) = m_\chi^2 \chi,
    \label{eq:MassiveKGEScalar}
\end{equation}
with the acoustic metric \eqref{eq:AcousticMetricSpinWaves} and the mass
\begin{equation}
    m_\chi^2(t) = \frac{2M}{\hbar^2} q(t) c_\mathrm{s,1}^2(t) = \frac{q(t)}{\hbar^2} (2 \bar n c_1(t) + q(t))
\end{equation}
Note that the finite magnetic field dresses the spin-exchange collisions with the quadratic Zeeman coefficient $q$, such that the spin waves now propagate with
\begin{equation}
    c_\mathrm{s,1}^2(t) = \frac{2 \bar n c_1(t) + q(t)}{2M}.
\end{equation}
For the transversal degree of freedom, we find the Proca equation
\begin{equation}
    \frac{1}{\sqrt{g_1}} \partial_\mu \left(\sqrt{g_1} g_1^{\mu \nu} \partial_\nu A^\sigma \right) = m_A^2 A^\sigma,
    \label{eq:ProcaEquation}
\end{equation}
with the mass
\begin{equation}
    m_A^2(t) = \frac{2M}{\hbar^2} q(t) c_\mathrm{s,1}^2(t) = \frac{q(t)}{\hbar^2} (2 \bar n c_1(t) + q(t)).
    \label{eq:ProcaMass}
\end{equation}
The Proca field can be defined as a massive, relativistic, spin-one field \cite{Heisenberg2019}, which is also known as the dark photon \cite{Fabbrichesi2020} (with its mass and thus its propagating degrees of freedom distinguishing it from the Maxwell field).
It has relevance in dark matter physics \cite{Graham2016,DeFelice2025} and modified gravity \cite{Heisenberg2019}.
An introduction to the Proca field on a flat spacetime can be found in \cite{MandlShaw2010}.
In \cref{eq:ProcaEquation}, the Proca field is minimally coupled to a curved spacetime described by the metric $g_1$ 
in the sense that its coupling to geometrical quantities only occurs via its Lagrangian density
\begin{equation}
        \mathcal{L} = - \frac{1}{4} F_{\mu \nu} F^{\mu \nu} - \frac{1}{2} m_A^2 g_1^{\mu \nu}  A_\mu A_\nu
        \label{eq:ProcaLagrangian}
\end{equation} 
being multiplied to the canonical volume form in the effective action $\Gamma = \int \mathrm{d}^{D+1} x \sqrt{g_1} \mathcal{L}$. 
A derivation of the field equation from the Lagrangian density \eqref{eq:ProcaLagrangian} can be found in Appendix \ref{sec:ProcaField}. 
Non-minimal coupling terms, on the other hand, would appear directly in the Lagrangian density \eqref{eq:ProcaLagrangian} and be of the form $\xi_1 R A_\mu A^\mu$ or $\xi_2 R_{\mu \nu} A^\mu A^\nu$, 
where $R_{\mu \nu}$ is the Ricci tensor, $R = g_1^{\mu \nu} R_{\mu \nu}$ its trace, and $\xi_1$ and $\xi_2$ are coupling constants.

The massive, relativistic form of the magnon dispersion relation has been found previously by Lamacraft \cite{Lamacraft2007}; however, the formulation as a Proca theory minimally coupled to a curved background is not present there.

In summary, the transversal degrees of freedom of the magnons (i.e.\@ the NGBs of SU$(2)$-symmetry-breaking) are analogue to a relativistic spin-one field $A_\mu$ which acquires a mass through the explicit SU$(2)$-symmetry-breaking for $q \neq 0$.
A time-dependent quadratic Zeeman coefficient $q$ can then be utilized to simulate a dynamic spacetime for the spin-one field.

\subsection{Ferromagnetic phase}
The ferromagnetic phase is naturally favoured by $p \neq 0$ and a small enough quadratic Zeeman-coupling $q < \bar n c_1$.
It can even occur for anti-ferromagnetic interactions ($c_1 > 0$) if  $p > \bar n c_1$.
For sufficiently positive $q \geq \bar n c_1$, the linear Zeeman shift needs to surpass $p > q/2$ in order to maintain this phase.  
In the following, we again consider only the positively polarized case where now then ground-state parameter is explicitly given by $\xi = \tfrac{1}{2}(\mathds{1}_2 + \sigma_z)$.
The U$(1)$-NGB is still described by the massless relativistic scalar field theory \eqref{eq:PhononActionFerro} with the acoustic metric \eqref{eq:AcousticMetricFerro}.

Since the external magnetic field leads to \emph{explicit} SU$(2)$-symmetry breaking, the rotational modes of the spin-density (or equivalently the nematic director) \eqref{eq:TransversalSpinDensityFlucFerro} now induce fluctuations of the effective potential of the form
\begin{equation}
  \frac{p-q}{2} \left((\phi_0^\mathrm{re})^2 + (\phi_0^\mathrm{im})^2 \right) \subset  \delta U + V_\mathrm{ext} \delta n^{(2)} 
\end{equation}
such that integrating out $\phi_0^\mathrm{re}$ for small momenta $E_k \ll (p-q)$, leads to 
\begin{equation}
    \Gamma_2^0 = - \frac{\hbar^2}{2} \int_{t,\bm{r}} \sqrt{g_0} \,  (g_0^{\mu \nu} \partial_\mu \phi_0 \partial_\nu \phi_0 + m_\phi^2 \phi_0^2),
\end{equation} 
where $\phi_0 \equiv \phi_0^\mathrm{im}/\sqrt{2M}$ with the acoustic metric
\begin{equation}
    (g_0)_{\mu \nu} = \begin{pmatrix} -1 & 0 \\ 0 & c_{\mathrm{s},0}^{-2} \, \delta_{ij} \end{pmatrix}, \qquad c_{\mathrm{s},0}^2(t) = \frac{p(t) - q(t)}{2M}.
\end{equation}
and the relativistic field mass
\begin{equation}
    m_\phi^2(t) = \frac{4M^2}{\hbar^2} c_\mathrm{s,0}^4(t) = \frac{1}{\hbar^2} (p(t) - q(t))^2.
\end{equation}
Similarly to \cref{subsubsec:NemSecNR}, we find that this theory must be considered in the non-relativistic limit where
\begin{equation}
    \hbar \omega_0(k) = \sqrt{(\hbar k c_0)^2 + (p-q)^2} \approx m_0 c_0^2 + \frac{\hbar^2 k^2}{2m_0} 
\end{equation}
with the heavy mass-term $m_0 c_\mathrm{s,0}^2 = p-q$ which is consistent with $m_\phi = m_0 c_\mathrm{s,0}^2/\hbar^2$.

For the nematic deformation mode ($\phi_-$), we find the same form of the effective action as in \cref{eq:EA_NematicDeformFerro}, but with the mass $m_- = 2(p - \bar n c_1)$.

\subsection{Anti-Ferromagnetic phase}
The anti-ferromagnetic phase is supported by $c_1 >0$ and $q<0$ whereas the spin-exchange interactions compete with the linear Zeeman effect $\bar n c_1 \geq \abs{p}$, and a positive quadratic Zeeman coupling $\bar n c_1 \geq q$. 
It is described by the order parameter
\begin{equation}
    \xi_\pm  =  \mathrm{e}^{\mathrm{i} \chi_\pm} \sqrt{(1 \pm f_z)/2}
    \label{eq:AFOrderParam}
\end{equation}
where $f_z \equiv p / \bar n c_1 \in [-1,1]$ and the phases are locked into $\chi_+ + \chi_- = \pi$ to minimize the effective potential \eqref{eq:effectivePotential} \cite{Jacob2012}. 
The background magnetization reads $\bar{\bm{F}} = \bar n f_z \bm{e}_z$,
such that one has magnetic order aligned with the external magnetic field if $f_z \neq 0$.
In addition to this dipole-moment, the background also carries a quadrupole moment that is described by the tensor
\begin{equation}
        \bar{\mathcal{Q}} / \bar n = (-\tfrac{1}{3} + \mathcal{A})  \, \bm{e}_\rho \otimes \bm{e}_\rho + (-\tfrac{1}{3} - \mathcal{A})   \, \bm{e}_\chi \otimes \bm{e}_\chi + \tfrac{2}{3} \bm{e}_z \otimes \bm{e}_z,
    \label{eq:BiAxialNematicOrder}
\end{equation}
with the radial and azimutal unit vectors in the xy-plane 
\begin{align}
    \bm{e}_\rho &= \cos\left(\delta \chi/2\right) \bm{e}_x + \sin\left(\delta \chi/2 \right) \bm{e}_y, \label{eq:eRho} \\
    \bm{e}_\chi &= -\sin \left(\delta \chi/2 \right) \bm{e}_x + \cos\left(\delta \chi/2 \right) \bm{e}_y \label{eq:eChi}, 
\end{align}
where we abbreviate the relative phase $\delta \chi = \chi_+ - \chi_- $ and have introduced the alignment parameter $\mathcal{A} = \sqrt{1-f_z^2}$ that quantifies the anisotropy of spin-fluctuations in that plane \cite{Zibold2016,deGennes1993}.
Note that the state at hand is \emph{not} a bi-axial nematic state which would be characterized by two mutually orthogonal, major and minor nematic directors associated to a rectangular symmetry described by the Klein four-group D$_4$ \cite{deGennes1993,Madsen2004,Mottram2014}. 
Since quantum spin-nematic states require time-reversal symmetry \cite{Chojnacki2024} one has to take $f_z = 0$ in that case, resulting in maximal alignment between the two axes ($\mathcal{A} = 1$), such that the nematic tensor takes the form 
\begin{equation}
    \bar{\mathcal{Q}} = (-2 \bar n) (\bm{e}_\chi \otimes \bm{e}_\chi - \tfrac{1}{3} \mathds{1}_3) \quad (\mathrm{for}\,  f_z = 0)
\end{equation}
which describes a uni-axial state with director $\bm{e}_\chi$ that is confined to the xy-plane.

The symmetry breaking pattern in the anti-ferromagnetic phase is
\begin{equation}
    \begin{aligned}
        &\mathrm{SU}(2)\xrightarrow{\mathrm{expl}.} \mathrm{SO}(2) \rtimes \mathbb{Z}_2 \xrightarrow{\mathrm{spont}.} \mathbb{Z}_2 \quad (f_z = 0), \\ 
        &\mathrm{SU}(2)\xrightarrow{\mathrm{expl}.} \mathrm{SO}(2) \xrightarrow{\mathrm{spont}.} \mathbb{Z}_2 \hspace{1.2cm}(f_z \neq 0).
    \end{aligned}
\end{equation}
where the spontaneous symmetry breaking is associated to the ground-state degeneracy parametrized by the two phases $\chi_\pm$ in \cref{eq:AFOrderParam} that fix the orientation of the nematic director $\bm{e}_\chi$ in the xy-plane (cref.\@ \cref{eq:eChi}).  
This results in the existence of two massless Nambu-Goldstone bosons as we shall see.
In analogy to the easy-plane ferromagnetic state, the non-magnetized state $f_z = 0$ has been described as an easy-plane polar state (EPP) \cite{Jacob2012}. 
We already discussed the opposite case of a nematic director pointing into the z-direction when analyzing the polar phase that is accordingly also referred to as the easy-axis polar phase (EAP) \cite{Liu2020}.
The second-order quantum phase-transition between these two regimes has been observed experimentally in \cite{Jacob2012}.

One finds the effective potential for the fluctuations 
\begin{equation}
        \begin{aligned}
            &\delta U + V_\mathrm{ext} \delta n^{(2)} \\[2pt]
            &= \frac{c_0 }{2} (\delta n^{(1)})^2 + \frac{c_1 }{2} \langle \delta \bm{F}^{(1)}, \delta \bm{F}^{(1)}  \rangle
            + \frac{q}{2}(\delta \mathcal{Q}_{zz}^{(2)} - \tfrac{2}{3}  \delta n^{(2)}),
        \end{aligned}
    \end{equation}
under use of the Thomas-Fermi relation 
\begin{equation}
    q + \bar n c_0 + V_\mathrm{ext}  =  \mu.
    \label{eq:eom-anti}
\end{equation}
Integrating out $\phi_0^\mathrm{re}$ results in the
\begin{equation}
\Gamma_2^0 = - \frac{\hbar^2}{2} \int \diff t\diff^2 \bm{r} \sqrt{g_0} \left(g_0^{\mu\nu} \partial_\mu \phi_0 \partial_\nu \phi_0 + m_\phi^2 \phi_0^2 \right),
\label{eq:AFScalarAction}
\end{equation}
as an effective action for  $\phi_0 = \phi_0^\mathrm{im}/\sqrt{2M}$ where one has the acoustic metric
\begin{equation}
    (g_0)_{\mu \nu} = \begin{pmatrix} -1 & 0 \\ 0 &  c_{\mathrm{s},0}^{-2} \delta_{ij} \end{pmatrix}, \qquad c_{\mathrm{s},0}^2 = \frac{q_- - q}{2M}.
\end{equation}
and the mass 
\begin{equation}
    m_\phi^2 = (q_+ - q) (q_- - q) / \hbar^2 = \frac{2M}{\hbar^2} (q_+ - q) c_\mathrm{s,0}^2
    \label{eq:MassAF}
\end{equation}
which notably vanishes at $ q_\pm \equiv \bar n c_1 (1 \pm \sqrt{1-f_z^2})$.
In particular, $q = q_-$ (where also $c_\mathrm{s,0}$ vanishes) defines the second-order boundary between the anti-ferromagnetic ($q < q_-$) and the easy-axis polar ($q \geq q_-$) phase \cite{Jacob2012}. 
Considering the dispersion relation $\hbar^2 \omega_0^2(k) = (E_k + q_+ - q) (E_k + q_- - q)$ one again finds that a quadratic form in the infrared regime $E_k \ll (q_+ - q)$ such that the validity of the emergent theory is confined to the non-relativistic limit. 
More concretely, we have 
\begin{equation}
    \Gamma_2 = \frac{\hbar^2}{2} \int_{\omega,\bm{k}} \frac{\omega^2 - \omega_0^2}{2M c_{\mathrm{s},0}^2} \phi_0^\mathrm{im}(-\omega,-\bm{k}) \phi_0^\mathrm{im}(\omega,\bm{k})
\end{equation}
with 
\begin{equation}
    \begin{aligned}
        \hbar \omega_0 = \sqrt{q- q_-} \sqrt{E_k + q_+ - q} \approx m_0 c_0^2+ \frac{\hbar^2 k^2}{2m_0},
    \end{aligned}
\end{equation}
at leading order. Here the rest-mass term is $$m_0 c_0^2 = \sqrt{(q_+ - q)(q_- - q)}$$ and related to the field mass up to a choice of units, $m_\phi = m_0 c_\mathrm{s,0}^2/\hbar^2.$
The scalar field of interest is related to spin-density and spin-nematic fluctuations via
\begin{equation}
    \begin{aligned}
    \delta F_x^{(1)} \pm \delta \mathcal{Q}_{xz}^{(1)} &= \sqrt{(1 \pm f_z) 2\bar n} \, \Re[\mathrm{e}^{-\mathrm{i} \chi_\pm} (\phi_0^\mathrm{re} + \mathrm{i} \phi_0^\mathrm{im})], \\
    \delta F_y^{(1)} \pm \delta \mathcal{Q}_{yz}^{(1)} &= \sqrt{(1 \pm f_z) 2\bar n} \, \Im[\mathrm{e}^{-\mathrm{i} \chi_\pm} (\phi_0^\mathrm{re} + \mathrm{i} \phi_0^\mathrm{im})]. 
    \end{aligned}
    \label{eq:ScalarModeAnti}
\end{equation}
For the remaining fluctuations, it is convenient to perform a U$(1)$-transformation, $\varphi_\pm \equiv \mathrm{e}^{- \mathrm{i} \chi_\pm} \phi_\pm$,
and to integrate out $\varphi_\pm^\mathrm{re}$ which results in
\begin{equation}
    \begin{aligned}
        \Gamma_2^\mathrm{\pm} &=  \frac{\hbar^2}{2}  \int_{t,\bm{r}}  \bigg \lbrace  \frac{c_0 + c_1}{ 4 \bar n c_0 c_1 (1-f_z)} (\partial_t \varphi_-^\mathrm{im})^2 - \frac{1}{2M} (\bm{\nabla} \varphi_-^\mathrm{im})^2 \\
        &+ \frac{c_0 + c_1}{4 \bar n c_0 c_1 (1+f_z) } (\partial_t \varphi_+^\mathrm{im})^2 - \frac{1}{2M} (\bm{\nabla} \varphi_+^\mathrm{im})^2  \\
        &- \frac{c_0 - c_1}{2 \bar n c_0 c_1 \sqrt{1-f_z^2} } (\partial_t \varphi_-^\mathrm{im}) (\partial_t \varphi_+^\mathrm{im}) \bigg \rbrace, 
    \end{aligned}
    \label{eq:SpinModeAnti}
\end{equation}
in the acoustic limit, $\hbar^2 \bm{\nabla}^2/(2M) \ll \bar n (c_0 + c_1) (1-f_z)$. 
For a homogeneous background density $\bar n$, and time-independent $c_0$ and $c_1$ we can diagonalize the effective action \eqref{eq:SpinModeAnti} in Fourier space and find 
\begin{equation}
    \begin{aligned}
        \Gamma_2^\mathrm{\pm} =  \frac{\hbar^2}{2} \int_{\omega,\bm{k}}  &\bigg \lbrace \psi_+(-\omega,-k)  \left(\frac{\omega^2}{c^2_{\mathrm{s},+}} -  k^2\right)  \psi_+(\omega,k)  \\
        &+ \psi_-(-\omega,-k) \left(\frac{\omega^2}{c^2_{\mathrm{s},-}}-  k^2\right) \psi_-(\omega,k) \bigg \rbrace \\
    \end{aligned}
    \label{eq:FinalGammaAF}
\end{equation}
where we introduced the fields
\begin{align}
        \psi_\pm = \frac{1}{\sqrt{2M}}\frac{1}{\sqrt{\mathcal{C}_\pm}} \left( \varphi_+^\mathrm{im} + W_\pm \varphi_-^\mathrm{im} \right),
\end{align}
with the mixing parameter
\begin{equation}
    W_\pm = \frac{(c_0 - c_1) \sqrt{1-f_z^2}}{(c_0 + c_1) f_z \pm \sqrt{(c_0 - c_1)^2 + 4 c_0 c_1 f_z^2}}
\end{equation}
and normalization
\begin{equation}
    \mathcal{C}_\pm = 1 \pm \frac{(c_0 + c_1)f_z}{\sqrt{(c_0 - c_1)^2 + 4c_0 c_1 f_z^2}}.
\end{equation}

\begin{figure*}
    \includegraphics[scale=0.4]{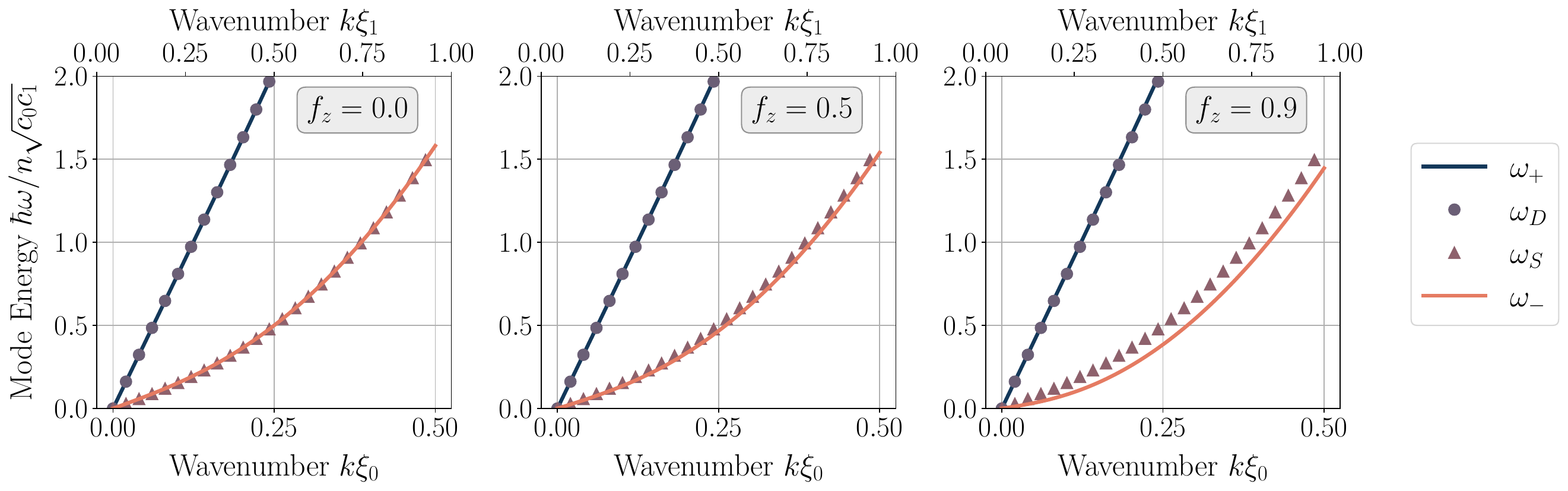}
    \caption{Dispersion relations $(\hbar \omega_\pm)^2$ of the exact eigenmodes formed by superpositions of phonons and magnons, resulting in an avoided crossing of pure density- and spin-modes with Bogoliubov dispersion $\hbar^2 \omega_\mathrm{D,S}^2$ for $c_0 / c_1 = 32$ which is representative of an antiferromagnetic $^{23}\mathrm{Na}$-BEC (cref.\@ \cref{tab:ExperimentalValues}).
    Due to the high ratio $c_0/c_1$, an external magnetic field $f_z \neq 0$ only marginally influences the dispersion relations of the eigenmodes. Perceptually uniform colormaps designed by \cite{Crameri2023} are used.}
    \label{fig:AvoidedCrossing}
\end{figure*}

In absence of an external magnetic field ($f_z = 0$), one has $W_\pm = \pm 1$ and $\mathcal{C}_\pm = 1$ such that the eigenbasis of the inverse propagator takes the form
\begin{equation}
    \psi_\pm = \frac{1}{\sqrt{2M}} \left(\varphi_+^\mathrm{im} \pm \varphi_-^\mathrm{im}\right),
\end{equation}
which can be identified as phase-fluctuations conjugate to
\begin{equation}
    \begin{aligned}
        \frac{1}{\sqrt{2}} (\varphi_+^\mathrm{re} + \varphi_-^\mathrm{re}) = \frac{\delta n^{(1)}}{\bar n}, \quad 
        \frac{1}{\sqrt{2}} (\varphi_+^\mathrm{re} - \varphi_-^\mathrm{re}) =  \frac{\delta F_z^{(1)}}{\bar n},
    \end{aligned}
\end{equation}
and represent the NGBs associated to spontaneous symmetry breaking of $\mathrm{U}(1)$ and $\mathrm{SO}(2)$, respectively.
Consequently, in presence of a magnetic field ($f_z \neq 0$) the eigenmodes are coupled phonon-magnon-excitations \cite{Kawaguchi2012} that obey the linear dispersion relation $\hbar \omega_\pm = c_{\mathrm{s},\pm} k$ with the sound-speeds
\begin{equation}
   M c_{\mathrm{s},\pm}^2 = \frac{\bar n}{2} \left[c_0 + c_1 \pm \sqrt{(c_0 - c_1)^2 + 4 f_z^2 c_0 c_1}\right]^{1/2}.
   \label{eq:csEffAF}
\end{equation}
where the discriminant is non-negative since $\lvert f_z \rvert \leq 1$.
Note that the magnon is represented by a single scalar degree of freedom here, since its associated spin-fluctuation needs to be perpendicular to the entire xy-plane to which the nematic director is confined.
This contrasts the vector properties of the magnon in the easy-plane polar phase where the nematic director possesses a two-dimensional orthogonal complement.
As shown in \cref{sec:ExplicitSymmetryBreakingDetailed}, the dispersion relation beyond the acoustic regime can be written as 
\begin{equation}
    \begin{aligned}
        \hbar^2 \omega_\pm^2(k) &= \frac{\hbar^2}{2}(\omega_\mathrm{D,0}^2(k) + \omega_\mathrm{S,0}^2(k)) \\
       &\pm \sqrt{\frac{\hbar^4}{4}\left(\omega_\mathrm{D,0}^2(k) - \omega_\mathrm{S,0}^2(k)\right)^2 + 4 f_z^2 E_k^2 \bar n^2 c_0 c_1},
    \end{aligned}
\end{equation}
which is an avoided crossing of unperturbed density- and longitudinal spin-modes (which occur for $f_z = 0$) and obey the Bogoliubov dispersion relations $\hbar^2 \omega_{\mathrm{D,0}}^2 = E_k(E_k + 2 \bar n c_{0})$ and $\hbar^2 \omega_{\mathrm{S,0}}^2 = E_k(E_k + 2 \bar n c_{1})$ in terms of the atomic kinetic energy $E_k = \hbar^2 k^2/2M$ (consider \cref{fig:AvoidedCrossing} for a visualization).

\section{Experimental prospects}
\label{sec:ExpFeasible}
In this final section we discuss the capability to observe the non-equilibrium effects described in this work with an emphasis on the emergent, curved Proca theory.
We base our analysis on common experimental values for the couplings of the effective theory which are collected in \cref{tab:ExperimentalValues}. 
Therein, we omit the linear Zeeman shift $p$ since it is oftentimes not given in experimental literature as it is fixed (for homogeneous magnetic fields) by global spin conservation 
\footnote{Equivalently, one can gauge-transform into the comoving spin-space that rotates with the p-induced Larmor frequency which separates from other energy scales in the spinor BEC \cite{Uhlmann2005,Stamper-Kurn2013,marti2014coherent}.}. 

Furthermore, we particularize to an emergent FLRW-spacetime by specifying a corresponding profile of the background density and the first and second Zeeman-couplings.
To preserve a certain stationary background density profile $\bar n(r)$ despite a variation of the interaction couplings $c_0(t),c_1(t)$ or the Zeeman coefficients $p(t),q(t)$,
one has to ensure that the Thomas-Fermi relation holds at all times.
For instance, in the context of phononic particle production via the time-dependent coupling $c_0(t)$, the external trapping potential $V_\mathrm{ext}$ has to be adjusted in time \cite{Tolosa2022,Viermann2022}.

\begin{table}
    \renewcommand{\arraystretch}{1.5} 
    \begin{tabular}{p{1.3cm} p{0.9cm} p{1.82cm} p{1.01cm} p{2.1cm} p{0.68cm}}
    \hline \hline
    Atom\footnotemark[1] & $ c_0/\lvert c_1 \rvert $ & $n \lvert c_1 \rvert / h$ [\SI{}{\hertz}] & $\xi_1$ [$\mu \mathrm{m}$] & $q/h$ [\SI{}{\hertz}]\footnotemark[2]&  Ref. \\ 
    \hline
    \multirow{4}{*}{$^{23}\mathrm{Na \,(af)}$}  & \multirow{4}{*}{32} 
                                                & 130          & 1.3 & $[-18, 2.5]_{\textrm{MW}}$ & \cite{Bookjans2011} \\ 
                                            &   & 100   & 1.5  & $[0, 70]_{\textrm{B}}$\footnotemark[3] & \cite{Stenger1998} \\ 
                                            &   & 38 & 2.4\footnotemark[3]  & $[4, 34]_{\textrm{B}}$ & \cite{Zibold2016}\\ 
                                            &   & 52, 48  & 16  & $[-240, 240]_{\textrm{MW}}$ & \cite{Zhao2014}\\ 
    \hline
    \multirow{4}{*}{$^{87}\mathrm{Rb \,(f)}$}   & \multirow{4}{*}{216}
                                                & 10\footnotemark[3]  & 2.4  & $[0.2, 280]_{\textrm{B}}$\footnotemark[2]     & \cite{Sadler2006} \\
                                            &   & 7.5\footnotemark[3] &  2.8\footnotemark[4]  &  $[0, 26]_{\textrm{B}}$\footnotemark[3]   & \cite{chang2005coherent} \\
                                            &   & 5.4 & 3.3\footnotemark[3] &$0.9_{\textrm{B}}$   & \cite{marti2014coherent} \\ 
                                            &   & 1.2 & 7.0\footnotemark[3] & $[-4, 2]_{\textrm{MW}}$\footnotemark[3]  & \cite{Pruefer2022} \\
    \hline
    $^7\mathrm{Li\, (f)}$     & 2 & 88 & $2.9$\footnotemark[3] & $[29.5, 610]_{\textrm{B}}$\footnotemark[3] &{\cite{huhStronglyFerromagnetic2020}}   \\ \hline \hline
    \end{tabular}
    \footnotetext[1]{(af) indicates anti-ferromagnetic interactions, $c_1 > 0$, whereas (f) indicates ferromagnetic interactions, $c_1 < 0$.}
    \footnotetext[2]{The subscript indicates whether a magnetic field (B) or microwave dressing (MW) is used to tune the quadratic Zeeman coefficient.}
    \footnotetext[3]{This value was inferred from other quantities or figures given in the reference. We use the convention $\xi_1 = \hbar/ \sqrt{2M \lvert c_1 n\rvert}$ for the spin healing length.}
    \caption{
    Realistic experimental values for the relevant paramters of the effective theory in spin-one Bose-Einstein condensates of common atomic species. 
    Values are rounded up to one decimal and their corresponding uncertainties can be obtained from the references.
    The values of $n c_1$ correspond to three-dimensional effective couplings $c_1$ and densities $n$. The lower-dimensional versions of $c_1$ can be computed in analogy to a scalar BEC \cite{Salasnich2002}.
    }
    \label{tab:ExperimentalValues}
\end{table}
\begin{table}
    \renewcommand{\arraystretch}{1.5} 
    \begin{tabular}{p{1.8cm} p{2.5cm} p{1cm}}
    \hline \hline
    Atom & Proca Mass [\SI{}{\hertz}] &  Ref. \\ 
    \hline
    \multirow{1}{*}{$^{23}\mathrm{Na}$}  &  $[141, 286]$ & \cite{Zhao2014}\\ 
    \hline
    \multirow{2}{*}{$^{87}\mathrm{Rb }$}   &  $[28, 290]$  & \cite{Sadler2006} \\
                                                &  $[21, 33]$   &  \cite{chang2005coherent}\\
    \hline
    $^7\mathrm{Li}$                       &  $[249, 692]$ &  \cite{huhStronglyFerromagnetic2020} \\ 
    \hline \hline
    \end{tabular}
    \caption{
    Estimates of the emergent Proca field mass \eqref{eq:ProcaMass} for various experimental realizations of spin-1 BECs in the polar ground state (with $q> 2\lvert c_1\rvert n$). Tuning the quadratic Zeeman-coefficient $q$ also changes the effective Proca mass.}
    \label{tab:ProcaMass}
\end{table}

\subsection{Simulation of cosmological Proca pair-production}

In the previous sections we derived an analogy between the transversal spin fluctuations of the polar phase and a Proca field subjected to a curved spacetime 
which emerges through spin-exchange collisions that proceed along the channel 
\begin{equation}
    2 \ket{m_F = 0} \longleftrightarrow \ket{m_F = +1} + \ket{m_F = -1},
    \label{eq:CollisionChannel}
\end{equation}
which is the only one allowed by total spin conservation. 
Typically anti-ferromagnetic interactions ($c_1 > 0$) and a positive quadratic Zeeman energy $q$ favor the left-hand-side whereas ferromagnetic interactions ($c_1 < 0$) and $q<0$ favor the right-hand-side \cite{Chang2004observation}.
For instance, the state where all condensate atoms occupy the $m_F = 0$ component (polar ground state) exhibits a dynamical unstable production of $(m_F = \pm 1)$-pairs driven by ferromagnetic spin interactions whose energy $2 n \lvert c_1 \rvert$ dominates the quadratic Zeeman energy $q$.
This behaviour initiates a quantum phase transition to a transversally magnetized state (easy-plane ferromagnetic phase, see lower green region of \cref{fig:PhaseDiagram}) which spontaneously breaks the SO(2)-symmetry of the polar ground state.
The instability was studied in many experiments where spin-textures and spin-domains that contain topolocical defects could be observed.
Furthermore, it was utilized to create spin-nematic (two-mode) squeezed vacuum states.

At this point we would like to mention that quenching to a region where the magnons become dynamically unstable would be analogue to a spacetime-signature-change towards euclidean signature,
which has been discussed for a scalar BEC that is quenched to the attractive regime and back to the repulsive regime \cite{Weinfurtner2007BoseNova} 
and has been experimentally realized in \cite{Chen2021} upon witnessing quasi-particle entanglement. These interesting aspects motivate to extend the present work to such an euclidean signature change.
In the following we contextualize guiding experimental studies with the Proca analogy.

\subsubsection{Implementation of temporal curvature}
\paragraph{Quench}
An experimentally feasible way to realize temporal curvature for the analogue Proca bosons consists of the well-known quantum-quench of the quadratic Zeeman coefficient $q$ via off-resonant microwave pulses (as explained in the context of \cref{eq:qmw}).
A quench of $q$ can be modeled by the Heaviside-profile \footnote{In fact, the microwave pulses utilized to quench $q$ actually have to be ramped up on a time-scale that typically is much faster than $h/n c_1$ in order to render them effectively instantaneous.}
\begin{equation}
    q(t) = q_\mathrm{i} \Theta(t_0 - t) + q_\mathrm{f} \Theta(t - t_0)
    \label{eq:Quenchq}
\end{equation} 
which, at spatially and temporally constant background density $\bar n$ and spin-exchange coupling $c_1$, corresponds to a spatially flat FLRW spacetime with an abruptly changing scale-factor of the form
\begin{equation}
    a_1^2(t) = \frac{2M}{q(t) + 2 \bar n c_1}.
    \label{eq:MagnonicScaleFactor}
\end{equation}
\paragraph{Ramp}
Alternatively, if one aims at a temporal profile of $a_1(t)$ that shall correspond to an expansion or contraction for a finite amount of time, 
one could apply a magnetic field ramp of the form
\begin{equation}
    B(t) = \frac{\sqrt{\Delta E_\text{hf}}}{g \mu_B} \left( \frac{2M}{a_1(t)^2}- 2 \bar n c_1\right)^{1/2}
    \label{eq:MagFieldRampProca}
\end{equation}
to tune the quadratic Zeeman coefficient \eqref{eq:quadraticZeemann}. Similar formulas could be derived by varying the spin-exchange coupling $c_1(t)$ and keeping $q$ constant in time.

A speciality of magnonic particle production is that at mean-field level the background state remains stationary despite the dynamics of $c_1(t)$ or $q(t)$, since the Thomas-Fermi relation \eqref{eq:ThomasFermiPolar} does not depend on these parameters.

\subsubsection{Relevant scales and Proca mass}
As can be deduced from \cref{tab:ExperimentalValues}, the typical energy scale of spin-dependent interactions $n \lvert c_1 \rvert$ can be up to two orders of magnitude smaller than the spin-independent one $n c_0$.
Thus, the spin-mixing dynamics are relatively slow compared to spin-independent contact interactions which is characterized by the timescale $h/n \lvert c_1 \rvert$ that may span over many milliseconds \cite{Stamper-Kurn2013}, such that \emph{in}-situ observations of the system typically proceed for hundreds of milliseconds \cite{Sadler2006,Pruefer2022}.
As a second consequence, the spin-healing length $\xi_1$ is large compared to the density healing length $\xi_0$, such that correlations are observed on relatively large scales.

Finally, the effective mass generated by the temporal variation of either the spin-coupling $c_1$ or the quadratic Zeeman effect $q$ needs to be sufficiently larger than the Proca mass (consider \cref{tab:ProcaMass} for realistic values). 
Typical effective masses generated in scalar BECs can be estimated by the following back-of-the-envelope calculation: In \cite{Schmidt2024}, analogue particle production was explained in terms of a scattering analogy where an effective scattering potential $V$ plays the role of a cosmological effective mass.
As can be deduced from the considerations in \cite{Schmidt2024}, if the field carries a mass $m^2$, one has to compare $V c_\mathrm{s}^2$ to $m^2$ where $c_\mathrm{s}$ is the speed of sound.
The companion experimental work \cite{Sparn2024} gave typical values for hallmark scattering potentials to be of the order $10^{-2} \mu\mathrm{m}^{-2}$. If we multiply this with typical sound speeds in scalar BECs, which are of the order $c_\mathrm{s} \simeq 1 \mu\mathrm{m}/\mathrm{ms}$, we get effective masses of the order of $10 \mathrm{kHz}$.
If the spinor couplings can be varied on similar time-scales as in scalar BECs by similar orders of magnitude, we can compare the value of $10 \mathrm{kHz}$ to the values in the table and conclude that the mass will play a relevant role in the particle production process but will not dominate it such that the signal should be strong enough.

\subsubsection{Observability of modes and two-point correlations}
The Heidelberg group \cite{Pruefer2022} succesfully probed the propagation of the three Bogoliubov modes of the easy-plane ferromagnetic phase of $^{87}\mathrm{Rb}$ by tracking a locally applied perturbation.
The spinor Goldstone mode, which is associated to the transversal spin-orientation, was found to propagate at approximately $0.1 \, \mu\mathrm{m}/\mathrm{ms}$ whereas the density mode (U(1)-Goldstone boson) was ten times faster in accordance with the ratio $\sqrt{c_0/c_1}$. 
To characterize thermalization and quantify the vacuum noise level, it was possible to extract the Fourier-spectrum of two-point correlations (structure-factor) of both the longitudinal and transverse spin components as well as all density components by combining a Stern-Gerlach sequence with time-of-flight measurements.
Quantitative agreement with theoretical continuum field-theory methods was found.
The spinor structure factors (which notably could be resolved in the acoustic EFT-regime $k\xi_1 > 1$) would be used as analogue particle spectra that are used to quantify pair-production of analogue relativistic vector particles from a thermal or vacuum state. 

\subsubsection{Structure formation via quenches}
In a pioneering experiment, the Berkeley group \cite{Sadler2006} quenched a two-dimensional $^{87}\mathrm{Rb}$ spinor gas from the polar ground state at a high value of $q_\mathrm{i} > 2 n \lvert c_1 \rvert$ to $q_\mathrm{f} < 2 n \lvert c_1 \rvert$ to the easy-axis ferromagnetic phase (cref.\@ \cref{fig:PhaseDiagram}).
The dynamical instability associated to the quantum-phase transition was characterized by an exponential growth of transverse magnonic correlations at a rate of $\tau = 15(4)$ ms which is in line with the spin-interaction energy $n \lvert c_1 \rvert / h = 8 \, \mathrm{Hz}$.
The resulting structures were transversely-magnetized spin-domains of typical size $10 \mu\mathrm{m}$ that contained spin-vortices of size $3 \mu \mathrm{m}$ as expected from the spin-healing length $\xi_1 = 2.4 \mu \mathrm{m}$ (consider \cite{Saito2007a,Saito2007b,Uhlmann2007,Damski2007} for a detailed analysis of the geometry and topology of these structures).

This experiment motivated a theoretical study \cite{Lamacraft2007} that analyzed the quantum-quench within pure quantum theory where deep ($q_\mathrm{f} \ll 2 n \lvert c_1 \rvert$) and shallow quenches ($n \lvert c_1 \rvert < q_\mathrm{f} < 2n \lvert c_1 \rvert$) were distinguished.
It was found that shallow quenches, whose final point lies in proximity of the polar phase, produce a characteristic light-cone propagation of two-point correlations at spin-wave velocity $c_\mathrm{s,1}$ which is in line with the literature on quasi-particle correlations and its associated distribution of entanglement \cite{Bravyi2006,Calabrese2006}.
Since the spin-exchange collisions are coherent \cite{chang2005coherent,marti2014coherent,Stamper-Kurn2013}, we expect the produced magnon pair to exhibit post-transition coherent interference effects at twice the magnon frequency in correspondence to the coherent oscillations of phononic quasiparticles that were observed in scalar BECs \cite{Hung2013,Chen2021,Viermann2022,Sparn2024}.

A quench from the polar to the easy-axis ferromagnetic phase was also studied in a structure formation experiment by the Heidelberg group \cite{Siovitz2025} with a particular emphasis on the predictive power of a double sine-gordon model as an low-energy effective field theory for the spinor phase.
Although the expansion point in field space as well as the expansion scheme are different than from this work, it will be interest to compare the found results.
In particular the theory developed in \cite{Siovitz2025} could be a resummation of terms that occured in our Bogoliubov approach.

While the two described experiments focussed on ferromagnetic systems, the dynamically unstable production of $(m_F = \pm 1)$-pairs from the $m_F = 0$ polar background has also been investigated via quenching an anti-ferromagnetic system from the polar to the anti-ferromagnetic phase \cite{Bookjans2011}. 
The instability rate was about $0.5/\mathrm{ms}$ for a shallow quench, it was found that a deep quench resulted in a pair-formation process that terminated after already $30 \mathrm{ms}$.
While the former $(\lvert q_\mathrm{f} \rvert \ll 1)$ requires a beyond-mean-field analysis due to condensate fragmentation, the experimental results corresponding to more negative values of $q_\mathrm{f}$ could be explained within a quantum rotor model (which can be derived from mean-field theory).

\subsubsection{Spin-nematic (two-mode) squeezing}
Quenching a ferromagnetic BEC prepared in the polar ground state across the quantum critical point $\lvert q \rvert = 2 n \lvert c_1 \rvert$ enabled the Atlanta group \cite{Hamley2012} to create a strongly spin-nematic squeezed state. 
The conjugate variables whose quadratures where measured to quantify the squeezing were $F_x$ and $\mathcal{Q}_{yz}$ (the conjugate variables $F_y$ and $\mathcal{Q}_{xz}$ are equally valid as \cite{Hamley2012}).
In an Heidelberg experiment \cite{Kunkel2022} these spin-nematic squeezed states were created and their entanglement structure was spatially resolved via full-state tomography. 
Therein both conjugate variables were combined in a single complex quantum field whose vacuum state was squeezed via off-resonant microwave dressing of $q$ which resulted in the creation of ($m_F = \pm 1$)-pairs.
In the following we want to make the case that this spin-nematic squeezing is related to two-mode squeezing in the phase space spanned by the relativistic Proca field and its conjugate momentum.
Let us first recall that Proca field $A_\mu$ was obtained as the divergence-free part of $\mathrm{Re} \bm{\varphi}_\perp/\sqrt{2M}$ which was introduced in \cref{sec:PolarPhaseSym,sec:PolarPhase} (where the mass factor establishes a relativistic mass dimension).
The latter can be related to the nematic fluctuations via
\begin{equation}
    \mathrm{Re} \, \bm{\varphi}_\perp = - \frac{1}{\sqrt{2\bar n}} \delta \mathcal{Q}_{xz} \bm{e}_x - \frac{1}{\sqrt{2\bar n}} \delta \mathcal{Q}_{yz} \bm{e}_y,
\end{equation}  
if we choose the nematic director aligned with the z-axis, as it can be shown by explicitly evaluating \cref{eq:VectorFieldDecomposition} in terms of the spin-nematic fluctuations calculated in \cref{sec:physFluc}.
The conjugate momentum of $A_\mu$ can be identified as the divergence-free component of $\hbar \, \mathrm{Im} \, \bm{\varphi}_\perp/\sqrt{2M}$.
Indeed, considering the on-shell relation between both fields, which expliciy derived in \cref{sec:ExplicitSymmetryBreakingDetailed}, one finds
\begin{equation}
    \hbar \, \mathrm{Im} \, \bm{\varphi}_\perp / \sqrt{2M} = \hbar^2 \sqrt{g_1} \partial_t \mathrm{Re} \, \bm{\varphi}_\perp /\sqrt{2M},
\end{equation}
in the acoustic limit where $g_1 = c_\mathrm{s,1}^{-4}$ is the determinant acoustic metric \eqref{eq:AcousticMetricSpinWaves}.
Note that the right-hand-side has the form of a relativistic momentum field conjugate to $\mathrm{Re} \, \bm{\varphi}_\perp /\sqrt{2M}$ 
which is minimally coupled to a spacetime with metric $g_1$ (consider \cite{Tolosa2022} for the corresponding expression in case of a scalar field).
Now, the imaginary part can be expressed via the spin-density fluctuations via 
\begin{equation}
    \mathrm{Im} \, \bm{\varphi}_\perp = \frac{1}{\sqrt{2 \bar n}} \delta F_y \bm{e}_x - \frac{1}{\sqrt{2 \bar n}} \delta F_x \bm{e}_y.
\end{equation}
Finally, we use that the commutator relations of the conjugate spin-nematic variables read \cite{Hamley2012,Kunkel2022}
\begin{equation}
    [\delta F_x, \delta Q_{yz}] = - 2 \mathrm{i} \bar n, \quad [\delta F_y, \delta \mathcal{Q}_{xz}] = 2 \mathrm{i} \bar n,
\end{equation}
such that one finds 
\begin{equation}
    [\mathrm{Re} \,(\varphi_\perp)_a, \hbar \, \mathrm{Im} \,(\varphi_\perp)_b ] = \mathrm{i} \hbar \delta_{ab},
\end{equation}
and similarly for their divergence-free components, respectively.
As a consequence, we conclude that spin-nematic squeezing between $\delta F_x$ and $\delta \mathcal{Q}_{xz}$ (or $\delta F_x$ and $\delta \mathcal{Q}_{yz}$) stemming from pair-creation of $m_F = \pm 1$ states via quenching the spin-dependent interactions that proceed along the collision channel \eqref{eq:CollisionChannel}
is related to two-mode squeezing between the relativistic Proca field $\mathrm{Re} \, \bm{\varphi}_\perp$ and its conjugate momentum $\hbar \, \mathrm{Im} \, \bm{\varphi}_\perp$, where the two-modes are counterpropagating momenta (stemming from the reality of $\mathrm{Re} \,(\bm{\varphi}_\perp)$ and $\mathrm{Im} \,(\bm{\varphi}_\perp$)) that carry an opposite magnetic quantum number $m_F = \pm 1$. 

\subsection{Acoustic FLRW-metric for scalar fields}

\subsubsection{Uni-metric cases: Polar and ferromagnetic phase}
The dynamics in the scalar sector of the magnonic fluctuations of the polar phase was shown to be analoge to a massive (or massless if $q=0$) Klein-Gordon equation \eqref{eq:MassiveKGEScalar}.
The background metric can be tuned similar to the Proca case as discussed in the context of \cref{eq:Quenchq,eq:MagnonicScaleFactor,eq:MagFieldRampProca}.

The massless QFTCS for the phonons of the polar and the ferromagnetic phase can be analyzed on similar grounds. 
For instance, in the ferromagnetic phase, the analogue line-element reads 
\begin{equation}
    \diff{s_+}^2 = -\diff{t}^2 + \frac{M}{\bar n(r) [c_0(t) + c_1(t)]} \diff^2 x.
\end{equation}
By choosing the density profile
\begin{equation}
    \bar n(r) = n_0 \left(1 \mp \frac{r^2}{R^2}\right)^2,
    \label{eq:ferroBackgroundDensity}
\end{equation}
one finds 
\begin{gather}
    \mathrm{d}s_+^2 = -\diff{t}^2 + a_+^2(t) \left(\frac{\diff{u}^2}{1 - \kappa u^2} + u^2 \diff \Omega^2 \right),
\end{gather}
with the spatial curvature scalar $\kappa = \mp 4/R^2$ and the scale-factor 
\begin{equation}
    a_+^2(t) = n_0 [c_0(t) + c_1(t)] / M.
\end{equation}
Note that this acoustic metric is similar to that of a scalar BEC \cite{Tolosa2022} 
but with an additional spin-dependent interaction term appearing (that is absent in the polar phase) 
since the density fluctuations are degenerate with longitudinal spin-fluctuations as the background is fully magnetized.

\subsubsection{Bi-metric cases}
The of the anti-ferromagnetic phase have a linear dispersion relation, one could map the effective action \eqref{eq:FinalGammaAF} to relativistic quantum field theories on bi-metric cosmological spacetimes, that are each associated to the analogue scale-factors $a_\pm(t) = 1/c_{\mathrm{s},\pm}(t)$, 
where the sound-speeds $c_\pm$ are defined in \cref{eq:csEffAF} and could be made time-dependent via tuning $c_0(t)$ and $c_1(t)$.
If $f_z =0$, the phonons and magnons separate such that one has the clear forms
\begin{equation}
    a_+^2(t) = n_0 c_0(t) / M \quad a_-^2 =  n_0 c_1(t) / M,
    \label{eq:BiMetricScaleFactors}
\end{equation}
assuming a spatially constant density profile $n = n_0$.
Here, the phononic spacetime would expand much faster since typically $c_0 \gg c_1$. 
As a further consequence of $c_0 \gg c_1$, the dependence of $c_\pm$ on $f_z$ is rather weak (cref.\@ \cref{fig:AvoidedCrossing}) such that the expressions \eqref{eq:BiMetricScaleFactors} could be used as an approximation for $f_z \neq 0$.
In \cite{Wang2025} a coupled bi-metric spacetime was found in a pseudo-spin-1/2-system, where the coupling was such that if one spacetime was expanding, the other one was contracting.
The situation here is different in the sense that the two spacetimes are independent of each other.

\section{Conclusion and Outlook}
\label{sec:conclusion}

We have shown the emergence of quantum field theories on effective spacetime geometries in terms of low-energy excitations of a spin-1 Bose-Einstein condensate.
The additional spin degrees of freedom can be utilized to investigate massive bosonic vector and scalar fields which are minimally coupled to curved acoustic spacetimes. 
In our analysis, we focused on stationary, mean-field ground states in the polar phase, the ferromagnetic phase and the anti-ferromagnetic phase and expanded in fluctuations in quadratic order. 
In each ground-state, the phase-fluctuation conjugate to the density mode is a Nambu-Goldstone boson with linear dispersion relation in the infrared that is associated to the spontaneous symmetry breaking of $\mathrm{U}(1)$, and can be cast into massless, relativistic theory, as for a scalar BEC \cite{Tolosa2022}.
The remaining fields of interest were identified in terms of density, spin-density, and spin-nematic fluctuations and classified according to their transformation behaviour under the little group of the corresponding ground state, as summarized in \cref{fig:PhaseDiagram,tab:Summary}. For the NGBs with a quadratic dispersion relation, the emergent theory represents the non-relativistic limit.

Perhaps most interestingly is the polar phase, which has three Nambu-Goldstone bosons with linear dispersion associated to the spontaneous symmetry breaking of $\mathrm{U}(1) \times\mathrm{SU}(2) \to \mathrm{SO}(2) \rtimes \mathbb{Z}_2$.
The NBGs associated to the spontaneous symmetry breaking of the spin group SU$(2)$ are spin-nematic excitations that rotate the nematic director. 
They transform as a vector under the little group $\mathrm{SO}(2) \rtimes \mathbb{Z}_2$ and are therefore described by a two-component vector field in the plane orthogonal to the magnetic field.
The transversal component in that plane could be cast into a relativistic Proca field that is minimally coupled to an acoustic geometry set by the polar background. The remaining scalar component behaves accordingly as a massive scalar field  (which both becomes massless if $q=0$).
In the current stage, the analogy only holds to a specific gauge choice of the Proca field. In future work we aim to investigate conditions for the description as an (emergent) gauge theory.
One motivation could be that the Proca field, also referred to as the dark photon, is a dark matter candidate \cite{Graham2016,Fabbrichesi2020,DeFelice2025}.

In the anti-ferromagnetic phase, the gauge symmetry group $\mathrm{U}(1) \times \mathrm{SU}(2)$ is explicitly broken down to $\mathrm{SO}(2)$ which is spontaneously broken by the order parameter.
This leads to two scalar Nambu-Goldstone bosons with linear dispersion whose superposition forms the exact eigenmodes of the system where an avoided crossing occurs in presence of an external magnetic field ($f_z \neq 0$).
Those NGBs are phase fluctuations conjugate to density and longitudinal spin-density fluctuations, respectively, and can be cast into massless, relativistic field theories on bi-metric spacetimes that are independent of each other. 

Finally, we propose to engineer the emergent acoustic spacetimes to a cosmological FLRW-form through radially symmetric trapping potentials and time-dependent magnetic fields. 
However, other spacetime geometries that may for example contain analogue horizons for spin waves are interesting configurations as well and warrant future study. 
For such scenarios, the present work can be extended to non-stationary ground states that carry a background current (like a superflowing spinor BEC).
Indeed, linearly dispersing magnonic excitations of a non-stationary background have already been proposed in \cite{Roldan-Molina2017} to simulate analogue Hawking radiation of massless, relativistic vector particles via a coupling to a background spin current from which an analogue event horizon emerges.

A central future application of our results lies in the quantum simulation of cosmological production for which proper experimentally observable spin-nematic correlators shall be defined.
Starting in the polar phase, analogue pair production of Proca quanta could be produced via magnetic field ramps that correspond to certain expansion or contraction scenarios or in a quantum quench of the quadratic Zeeman-coefficient via off-resonant microwave pulses resulting in spin-nematic squeezed states \cite{Hamley2012,Kunkel2019,Kunkel2022}.
A prime candidate would be a quench from the polar to the easy-plane ferromagnetic phase \cite{Lannig2023,Siovitz2025}, realizing production of $m_F = \pm 1$-pairs from $m_F= 0$ quasi-particle states.

We are optimistic that the magnonic correlations spread on time- and length-scales that can be observed with modern experimental techniques at a relatively high temporal resolution.
The capability to perform spatially resolved full-state tomography \cite{Kunkel2022} provides a promising avenue to study quantum entanglement of analogue relativistic vector particles via spin-nematic squeezing.

A final future avenue consists in analyzing the quantum phase transitions of the spinor BEC in light of the constructed cosmological analogues with regards to cosmological phase transitions or the Kibble-Zurek mechanism. 
Studies regarding the latter already exist in context of the spinor BECs \cite{Saito2007a,Saito2007b,Uhlmann2007,Damski2007}.
In total, we believe that spinor Bose-Einstein condensates provide manifold possibilities to study analogue cosmology or gravity in more general terms. 

\section*{Acknowledgements}
The authors thank H.Gies, L.Janssen, M.Oberthaler and Y.Deller for fruitful discussions. 
C.F.S acknowledges support by the Studienstiftung des Deutschen Volkes and the Deutsche Forschungsgemeinschaft (DFG) under Grant No 406116891 within the Research Training Group RTG 2522/1.
This research was funded in whole or in part by the Austrian Science Fund (FWF) [10.55776/FG5]. For open access purposes, the author has applied a CC BY public copyright license to any author-accepted manuscript version arising from this submission.

\appendix
\clearpage
\newpage

\section{SU(2)-transformation of spin multipoles}
\label{sec:SU2Transformation}
Let us consider an SU$(2)$ group element $U \in \mathrm{SU}(2)$, under which the field transforms as $\Phi_{AB} \to U_A^C U_B^D \Phi_{CD}$. Then the spin density (or dipole) transforms according to 
\begin{equation}
    F_i = \Phi^*_{AB} (\sigma_i)^{BC} \Phi_{CA} \to \Phi_{XY}^* (U^\dagger)_Y^B (\sigma_i)_B^C U_C^Z \Phi_{ZX},
\end{equation}
where we used unitarity, $(U^*)_B^A U_A^C = \delta_A^C$. 
The more compact form
\begin{equation}
    F_i \to \tr (\Phi^* U^\dagger \sigma_i U \Phi),
\end{equation}
highlights that $\bm{F}$ transforms as a vector under spin rotations.
This expression can be further simplified under use of the homomorphism between SO$(3)$ and SU$(2)$ \cite{Cornwell1997}
\begin{align}
        U^\dagger \sigma_i U &= \mathcal{R}_{ij} \sigma_j, \label{eq:TransformationSUtoSO}  \\ 
        \mathcal{R}_{ij} &\equiv \tfrac{1}{2} \tr (\sigma_i U \sigma_j U^\dagger), \label{eq:RotationMatrixDef}
\end{align}
resulting in 
\begin{equation}
    F_i \to \mathcal{R}_{ij} F_j,
\end{equation}
i.e\@. internal spin rotations result in external space rotations of $\bm{F}$. 
Similarly, one finds
\begin{equation}
    \begin{aligned}
        &\tr(\Phi^* \sigma_i \Phi \sigma_j^T) \\
        &= \Phi^*_{AB} (\sigma_i)^{BC} \Phi_{CD} (\sigma_j)^{AD} \\
        &\to \Phi_{XY}^* (U^\dagger)_Y^B (\sigma_i)_B^C U_C^Z \Phi_{ZW} U_D^W (\sigma_j)_A^D (U^\dagger)_X^A, \\
    \end{aligned}
\end{equation}
which can be written as 
\begin{equation}
    \tr(\Phi^* \sigma_i  \Phi \sigma_j^T)   \to \tr(\Phi^* (U^\dagger \sigma_i U)  \Phi (U^\dagger \sigma_j U)^T),
\end{equation}
such that it becomes apparent that the spin-nematic tensor  (or quadrupole moment) transforms as a tensor,
\begin{equation}
    \mathcal{Q}_{ij} \to \mathcal{R}_{ik} \mathcal{R}_{jl} \mathcal{Q}_{kl}. 
\end{equation}

\section{Mean field ground states of a spin-one BEC}
\label{sec:groundstate}
In the following, we derive ground states which are investigated in the main text in the context of mean field theory. 
We omit the broken-axisymmetric phase, which is for example discussed in \cite{Kawaguchi2012}, as it is not relevant for the main text.
A necessary condition for the minima of the effective potential \eqref{eq:effectivePotential} is given by the multi-component Gross-Pitaevskii equation
\begin{equation}
    \begin{aligned}
        0 &= \frac{\delta \Gamma}{\delta \Phi_{AB}^*} \\
        &= \left( \mathrm{i} \hbar \partial_t + \frac{\hbar^2}{2M} \bm{\nabla}^2 - \frac{q}{2} + \tilde \mu \right) \Phi_{AB} \\[2pt]
        &\qquad + c_1 F_k (\sigma_k)_{BC} \Phi_{CA}  - p \, \Phi_{CA} (\sigma_3)_{BC} \\[2pt]
        & \qquad + \frac{q}{2} (\sigma_3)_{BC} \Phi_{CD} (\sigma_3)_{DA},
    \end{aligned}
\label{eq:GPESU2}
\end{equation}
restricted to stationary background solutions, $\partial_t \Phi_{AB} = 0$.
Note that we introduced $\tilde \mu = \mu - V_\mathrm{ext} - \bar n c_0$ for the sake of brevity. 
The stationary background solutions are parametrized via 
\begin{equation}
    \begin{aligned}
        \bar \Phi (r) &= \sqrt{\bar n(r)} \left( \mqty{ \xi_+ & \xi_0 / \sqrt{2} \\ \xi_0 / \sqrt{2} & \xi_- }\right). \\
    \end{aligned}
\end{equation} 
and contrained by the normalization
\begin{equation}
    \lvert \xi_0 \rvert^2  + \lvert \xi_- \rvert^2 + \lvert \xi_+ \rvert^2 = 1.
    \label{eq:BackgroundNormalization}
\end{equation}
Assuming a sufficiently smooth background density profile $\bar n(r)$ (which is reasonable for long-wavelength effective theories),
allows us to neglect quantum pressure terms and results in the following set of algebraic equations
    \begin{align}
        \left( - \tilde \mu - p + q + \bar n c_1 \bar f_z   \right) \xi_+ + \bar n c_1 (\xi_+ + \xi_-^*) \xi_0^2  &= 0, \label{eq:Spin1GPEPlus} \\[2pt]
    \left(- \tilde \mu + \bar n c_1 ( \lvert \xi_+ \rvert^2 + \lvert \xi_- \rvert^2 + 2 \xi_+ \xi_- ) \right) \xi_0 &= 0, \label{eq:Spin1GPE0} \\[2pt]
        \left( - \tilde \mu + p + q - \bar n c_1 \bar f_z   \right) \xi_- + \bar n c_1 (\xi_+^* + \xi_-) \xi_0^2  &= 0, \label{eq:Spin1GPEMinus}
    \end{align}
where we chose a global phase such that $\xi_0 \in \mathbb{R}$. 
Note that this set is equivalent to the one presented in \cite{Kawaguchi2012}. 
In the following, it will be advantageous to introduce the specific magnetization in the z-direction
\begin{equation}
    \bar f_z \equiv (\lvert \xi_+ \rvert^2 - \lvert \xi_- \rvert^2) = \bar F_z / \bar n. \label{eq:Fz}
\end{equation} 

\subsection*{Polar phase}
For $\xi_0 \neq 0$, we require that 
\begin{equation}
V_\mathrm{ext} - \mu + \bar n c_0 = - \bar n c_1 (\lvert \xi_+ \rvert^2 + \lvert \xi_- \rvert^2 + 2 \xi_+ \xi_- ) ,
\label{eq:PolarPhaseSolCondition}
\end{equation}
to fulfill \cref{eq:Spin1GPE0}. 
Then, we must have $\Im(\xi_+ \xi_-) = 0$ since the left-hand-side of \eqref{eq:PolarPhaseSolCondition} must be real.

Furthermore one can choose $\bar F_y = 0$ due to azimutal symmetry, such that 
\cref{eq:Spin1GPEMinus,eq:Spin1GPEPlus} can be brought into the form 
\begin{align}
   0 &=  (V_\mathrm{ext} - \mu + \bar n c_0 + q + 2 \bar n c_0 \xi_0^2) (\xi_+ + \xi_-) \nonumber \\
   &\quad - (p - \bar n c_1 \bar f_z) (\xi_+ - \xi_-), \label{eq:BrokenAxiNonTrivial1} \\
   0 &=  (V_\mathrm{ext} - \mu + \bar n c_0 + q) (\xi_+ - \xi_-) \nonumber \\
    &\quad - (p - \bar n c_1 \bar f_z) (\xi_+ + \xi_-) \label{eq:BrokenAxiNonTrivial2}.
\end{align}

The polar phase is then characterized by the parameter configuration
\begin{equation}
    \xi_0 = 1,\quad  \xi_\pm = 0, \quad \text{ with }\quad \left( V_\mathrm{ext} - \mu + \bar n c_0 \right) = 0,
\end{equation}
and carries no background magnetization, $\bm{\bar{F}} = 0$.
However, it still carries the magnetic energy $U_\mathrm{mag} = \bar n q$, due to quadrupolar order $Q_{zz} = 2 \bar n/3.$
This phase occurs when the quadratic Zeeman coefficient fulfills
\begin{equation}
    \begin{cases} 
     q > 2 \lvert c_1 \rvert \bar n & \text{for } c_1 < 0, \\
     q \geq 0 & \text{for } c_1 = 0, \\
     q > 0 & \text{for } c_1 > 0,
     \end{cases}
     \label{eq:PolarPhaseConditions}
 \end{equation}
 and $p=0$ can be taken without loss of generality due to magnetization conservation.

\subsection*{Ferromagnetic and antiferromagnetic phase}

\subsubsection*{Ferromagnetic phase}
\label{sec:FerroSol}
One possible solution to \cref{eq:Spin1GPE0} is given by $\xi_0 = 0$, for which the remaining equations can be solved by 
\begin{equation}
    \begin{aligned}
        \xi_+ = \mathrm{e}^{\mathrm{i} \chi_+}, \quad  \xi_- &= 0, \\
        V_\mathrm{ext} - \mu + \bar n c_0 - p + q + \bar n c_1  &= 0,
    \end{aligned}
    \label{eq:FerroSolPos}
\end{equation}
i.e.\@ the positively polarized case ($\bar f_z = 1$), or 
\begin{equation}
    \begin{aligned}
        \xi_+ = 0, \quad \xi_- &= \mathrm{e}^{\mathrm{i} \chi_-}, \\
         V_\mathrm{ext} - \mu + \bar n c_0 + p + q - \bar n c_1 &= 0,
    \end{aligned}
    \label{eq:FerroSolNeg}
\end{equation}
i.e.\@ the negatively polarized case ($\bar f_z = -1$).
Here, $\chi_\pm$ are arbitrary global phases. These states are referred to as ferromagnetic, since they entail $\bar f_z = \pm 1$. 
The minimized magnetic energy per particle is 
\begin{equation}
    \frac{U_\mathrm{mag}}{\bar n} = \frac{1}{2} \bar n c_1 \mp p + q, 
\end{equation}
where the upper sign corresponds to the positively polarized and the lower sign to the negatively polarized case. 

Consequently, ferromagnetic interactions ($c_1 < 0$), which tend to maximize the total magnetization, prefer this phase.
The linear Zeeman term supports this cause, as it favours a magnetization in the direction of the external magnetic field,
whereas the quadratic Zeeman term favours magnetized state if $q<0$; and a magnetically neutral state if $q>0$, which results in a competition between $p$ and $q$.  

Hence, for $c_1 \leq 0$, the ferromagnetic phase is constrained by the conditions
\begin{equation}
    \begin{cases}
    p > 0 & \text{if }  \quad q <0, \\
    p > q & \text{if } \quad  q \geq 0. \\
    \end{cases}
\end{equation}
If the linear Zeeman term becomes to weak (in absolute terms), a second-order phase transition into the broken-axisymmetric phase occurs for positive $q$.
For negative $q$, the system remains in the ferromagnetic phase unless $p=0$ where a first-order phase transition into the polar phase takes place.  

Interestingly, the ferromagnetic phase can also occur for anti-ferromagnetic interactions ($c_1 > 0$) if the linear Zeeman term is strong enough, i.e.\@
\begin{equation}
    \begin{cases} 
    p > c_1 \bar n & \text{if }  q <  \bar n c_1 / 2,  \\
    p > q + c_1 \bar n / 2 & \text{if } q \geq \bar n c_1 /2.
    \end{cases}
    \label{eq:FerroPhaseConditions}
\end{equation}
If the linear Zeeman shift becomes weaker than the anti-ferrogmagnetic interactions, $p <= \bar n c_1$, a second-order phase transition into the anti-ferromagnetic phase occurs,
for small enough $q < \bar n c_1/2$. If $q \geq \bar n c_1/2$, the ferromagnetic phase transitions into the polar phase with a first-order boundary. 

\subsubsection*{Antiferromagnetic phase}
The other solution for the case $\xi_0 = 0$ can be found by casting \cref{eq:Spin1GPEPlus,eq:Spin1GPEMinus} into the form
\begin{align}
    0 &= \left( V_\mathrm{ext} - \mu + \bar n c_0 + q \right) (\xi_+ - \xi_-) \nonumber \\
    &\quad - (p - \bar n c_1 \bar f_z) (\xi_+ + \xi_-), \label{eq:AFIntermediateUpper} \\[2pt]
   0 &= \left( V_\mathrm{ext} - \mu + \bar n c_0 + q \right) (\xi_+ + \xi_-) \nonumber \\
   &\quad - (p - \bar n c_1 \bar f_z) (\xi_+ - \xi_-) . \label{eq:AFIntermediateLower}
\end{align}
To fulfill the coherent normalization constraint \eqref{eq:BackgroundNormalization}, we search for non-trivial solutions of \cref{eq:AFIntermediateUpper,eq:AFIntermediateLower}. 
Hence, we require that
\begin{equation}
 \left( V_\mathrm{ext} - \mu + \bar n c_0 + q \right)^2 = (p - \bar n c_1 \bar f_z)^2 = 0,
\end{equation}
where the first equality follows from squaring both equations. 
Now, the spin components of interest follow from the normalization condition \eqref{eq:BackgroundNormalization} and read
\begin{equation}
    \xi_+ = \mathrm{e}^{\mathrm{i} \chi_+} \sqrt{\frac{1 + \bar f_z}{2}}, \quad \xi_- = \mathrm{e}^{\mathrm{i} \chi_-} \sqrt{\frac{1 - \bar f_z}{2}},
\end{equation}
where $\bar f_z = p/ \bar n c_1$ and $\chi_\pm$ are relative phases which are constrained by $\chi_ + + \chi_- = \pi$ to minimize the effective potential \cite{Jacob2012}.
This solution is called the antiferromagnetic phase as both the $m_F = +1$ and $m_F = -1$ components are occupied. 
The ground-state degeneracy stemming from the phases $\chi_\pm$ can be captured by the gauge transformation 
\begin{equation}
    \begin{aligned}
        &\left( \mqty{\sqrt{1 + f_z}  & 0 \\ 0 & \sqrt{1 - f_z} }  \right) \\[2pt]
        &= \mathrm{e}^{- \tfrac{\mathrm{i}}{2}(\chi_+ + \chi_-)} Z \left( \mqty{ \mathrm{e}^{\mathrm{i} \chi_+} \sqrt{1 + f_z}  & 0 \\ 0 & \mathrm{e}^{\mathrm{i} \chi_-} \sqrt{1 - f_z} }  \right) Z^\mathrm{T}
    \end{aligned}
\end{equation}
where $Z = \exp[- \tfrac{1}{4}\mathrm{i} (\chi_+ - \chi_-)  \sigma_z] $ is a spin-rotation by $(\chi_+ - \chi_-)/2$ around the z-axis.

Notably, for $f_z =0$, one finds a so-called easy-plane polar state that can be written as $\xi_\mathrm{EAP} = \sigma_x / \sqrt{2}$ without loss of generality (as discussed in the main text).
Interestingly it lies on the orbit of the (easy-axis) polar phase $\xi_\mathrm{EPP} = \mathds{1}_2 / \sqrt{2}$ under actions of the latter's little group $\mathrm{SO}(2)$.
More concretely, applying a $\pi/2$-spin-rotation around the x-axis followed by a U(1)-transformation, one finds
\begin{equation}
    \xi_\mathrm{EPP} = \mathrm{e}^{\mathrm{i}\pi/2} \exp(- \tfrac{1}{4}\mathrm{i} \pi  \sigma_x) \, \xi_\mathrm{EAP} \, \exp(- \tfrac{1}{4}\mathrm{i} \pi  \sigma_x)^\mathrm{T}.
\end{equation}

Since the magnetic energy per particle is
\begin{equation}
    \frac{U_\mathrm{mag}}{\bar n}= - \frac{p^2}{2 \bar n c_1} + q
\end{equation}
one has the phase boundaries 
\begin{equation}
    \begin{cases}
        \abs{p} \leq c_1 \bar n & \text{for } q < 0 \\
        \sqrt{2 c_1 \bar n q} \leq \abs{p} \leq c_1 \bar n  & \text{for } 0 \leq q \leq c_1 \bar n / 2. \\
    \end{cases}
\end{equation}
As a consequence, the anti-ferromagnetic phase can only occur for $c_1 > 0$, and only if the linear Zeeman shift is small enough (otherwise the system tends to the ferromagnetic state with a second-order boundary).
Similarly, if $q$ becomes too large, the system transitions into the polar phase with a first-order boundary.

\section{Perturbative scheme and identification of fluctuating fields}  
\label{sec:physFluc}
In the following, we identify fluctuations in the density via 
\begin{equation}
    n(\bar \Phi + \delta \Phi) \equiv \bar n + \delta n,
\end{equation}
the spin-density via 
\begin{equation}
F_i(\bar \Phi + \delta \Phi) \equiv \bar F_i + \delta F_i,
\end{equation}  
and the spin-nematic tensor via 
\begin{equation}
    \mathcal{Q}_{ij}(\bar \Phi + \delta \Phi) \equiv \bar{\mathcal{Q}}_{ij} + \delta \mathcal{Q}_{ij}.
\end{equation} 
To that end we insert $\Phi = \bar \Phi + \delta \Phi$ into \cref{eq:density,eq:spindensity,eq:nematictensor} and expand up to second order. 
This yields
\begin{equation}
    \bar n = \tr (\bar \Phi^* \bar \Phi), \quad \delta n = \delta n^{(1)} + \delta n^{(2)}, 
\end{equation}
where
\begin{equation}
    \begin{aligned}
        \delta n^{(1)} &= \tr (\delta \Phi^* \bar \Phi) + \tr (\bar \Phi^* \delta \Phi) \\
        \delta n^{(2)} &= \tr (\delta \Phi^* \delta \Phi).
    \end{aligned}
\end{equation}
Furthermore we have
\begin{equation}
    \bar F_i = \tr (\bar \Phi^* \sigma_i \bar \Phi), \quad \delta F_i = \delta F_i^{(1)} + \delta F_i^{(2)}, 
\end{equation}
where
\begin{equation}
    \begin{aligned}
        \delta F_i^{(1)} &= \tr (\delta \Phi^* \sigma_i \bar \Phi) + \tr (\bar \Phi^* \sigma_i \delta \Phi),  \\
        \delta F_i^{(2)} &= \tr (\delta \Phi^* \sigma_i \delta \Phi).
    \end{aligned}
\end{equation}
Finally, one finds 
\begin{equation}
    \bar{\mathcal{Q}}_{ij} = \tr (\bar \Phi^* \sigma_i \bar \Phi \sigma_j^T) - \tfrac{1}{3} \bar n \delta_{ij}, \quad \delta \mathcal{Q}_{ij} = \delta \bar{\mathcal{Q}}_{ij}^{(1)} + \delta \bar{\mathcal{Q}}_{ij}^{(2)}
\end{equation}
where
\begin{equation}
    \begin{aligned}
        \delta \mathcal{Q}_{ij}^{(1)} &= \tr (\delta \Phi^* \sigma_i \bar \Phi \sigma_j^T) + \tr (\bar \Phi^* \sigma_i \delta \Phi \sigma_j^T)  - \tfrac{1}{3} \delta n^{(1)}  \delta_{ij}, \\ 
        \delta \mathcal{Q}_{ij}^{(2)} &= \tr (\delta \Phi^* \sigma_i \delta \Phi \sigma_j^T) - \tfrac{1}{3} \delta n^{(2)}  \delta_{ij}. 
    \end{aligned}
\end{equation}
The linear fluctuations depend on the specific ground state configuration and shall be collected in the following.
The quadratic fluctuations are given thereafter. To evaluate the projections it is most convenient to represent the fluctuations in the basis of Pauli matrices, i.e.\@
\begin{equation}
    \delta \Phi = \frac{1}{\sqrt{8}} (\phi_+ + \phi_-) \mathds{1}_2 + \frac{1}{\sqrt{8}} (\phi_+ - \phi_-) \sigma_z + \frac{1}{2} \phi_0 \sigma_x.
\end{equation}

\subsection{Polar phase}
\label{sec:physFlucPolar}
Here, the density fluctuation reads
\begin{equation}
    \begin{aligned}
        \delta n^{(1)} &= \sqrt{2\bar n} \, \phi_0^\mathrm{re},
    \end{aligned}
\end{equation}
whereas the spin-density fluctuations are
\begin{align}
    \delta F_x^{(1)} &= \sqrt{\bar n} (\phi_-^\mathrm{re} + \phi_+^\mathrm{re} ), \\
    \delta F_y^{(1)} &=\sqrt{\bar n} (\phi_-^\mathrm{im} - \phi_+^\mathrm{im}), \\
    \delta F_z^{(1)} &= 0.
\end{align}

For the diagonal spin-nematic fluctuations we have 
    \begin{equation}
        \delta \mathcal{Q}_{ij}^{(1)} = - \frac{\delta n^{(1)}}{\bar n} \bar{\mathcal{Q}}_{ij} \quad  (\text{for } i=j), 
    \end{equation}
where
\begin{equation}
    \bar{\mathcal{Q}} = - 2 \bar n \,  \mathrm{diag}(1/3,1/3,-2/3)
\end{equation} 
is the nematic tensor of the polar background.
The offdiagonal fluctuations read
\begin{align}
    \delta \mathcal{Q}_{xy}^{(1)} &= 0, \\
    \delta \mathcal{Q}_{xz}^{(1)} &= - \sqrt{\bar n} \, (\phi_-^\mathrm{re} - \phi_+^\mathrm{re}), \\
    \delta \mathcal{Q}_{yz}^{(1)} &= - \sqrt{\bar n} \, (\phi_-^\mathrm{im} + \phi_+^\mathrm{im}).
\end{align}

\subsection{Ferromagnetic phase}
\label{sec:physFlucFerro}
Similar to the polar phase, the density fluctuation is 
\begin{align}
    \delta n^{(1)} &= \sqrt{2 \bar n} \, \phi_+^\mathrm{re},
\end{align}
For the spin-density fluctuations, one finds 
\begin{align}
    \delta F_x^{(1)} &= \sqrt{\bar n} \, \phi_0^\mathrm{re}, \\
    \delta F_y^{(1)} &= \sqrt{\bar n} \, \phi_0^\mathrm{im}, \\
    \delta F_z^{(1)} &= \sqrt{2\bar n} \, \phi_+^\mathrm{re}.
\end{align}
For the diagonal spin-nematic fluctuations we have 
\begin{equation}
    \delta \mathcal{Q}_{ij}^{(1)} = - \frac{\delta n^{(1)}}{\bar n} \bar{\mathcal{Q}}_{ij} + \sqrt{2 \bar n} \, (\phi_-^\mathrm{re} \delta_{ix} -  \phi_-^\mathrm{re} \delta_{iy})
\end{equation}
with the nematic tensor of the ferromagnetic background
\begin{equation}
    \bar{\mathcal{Q}} = \bar n \,  \mathrm{diag}(1/3,1/3,-2/3). 
\end{equation}
The off-diagonal terms are 
\begin{align}
    \delta \mathcal{Q}_{xy}^{(1)} =  \sqrt{2 \bar n} \, \phi_-^\mathrm{im}, \quad
    \delta \mathcal{Q}_{xz}^{(1)} = \sqrt{\bar n} \, \phi_0^\mathrm{re}, \quad
    \delta \mathcal{Q}_{yz}^{(1)} = \sqrt{\bar n} \, \phi_0^\mathrm{im}.
\end{align}

\subsection{Anti-Ferromagnetic phase}
\label{sec:physFlucAF}
Abbreviating 
\begin{align}
    \phi_+ &\equiv \phi_+^\mathrm{re} + \mathrm{i} \phi_+^\mathrm{im}, \\
    \phi_0 &\equiv \phi_0^\mathrm{re} + \mathrm{i} \phi_0^\mathrm{im}, \\
    \phi_- &\equiv \phi_-^\mathrm{re} + \mathrm{i} \phi_-^\mathrm{im},
\end{align}
we find
\begin{align} 
    \frac{\delta n^{(1)}}{\sqrt{\bar n}} &= \sqrt{1+f_z} \Re (\mathrm{e}^{-\mathrm{i} \chi_+} \phi_+) + \sqrt{1-f_z}  \Re( \mathrm{e}^{-\mathrm{i} \chi_-} \phi_-) .
    \label{eq:AFPhonon}
 \end{align}
for the density-fluctuation. 
The spin-density fluctuations are 
 \begin{align}
    \frac{\delta F_x^{(1)}}{\sqrt{\bar n/2}} &= \sqrt{1+f_z} \Re ( \mathrm{e}^{-\mathrm{i} \chi_+} \phi_0) + \sqrt{1-f_z} \Re ( \mathrm{e}^{-\mathrm{i} \chi_-} \phi_0), \\
    \frac{\delta F_y^{(1)}}{\sqrt{\bar n/2}} &= \sqrt{1+f_z} \Im ( \mathrm{e}^{-\mathrm{i} \chi_+} \phi_0) - \sqrt{1-f_z} \Im ( \mathrm{e}^{-\mathrm{i} \chi_-} \phi_0), \\
    \frac{\delta F_z^{(1)}}{\sqrt{\bar n}} &=  \sqrt{1 + f_z} \Re ( \mathrm{e}^{-\mathrm{i} \chi_+} \phi_+) - \sqrt{1-f_z} \Re ( \mathrm{e}^{-\mathrm{i} \chi_-} \phi_-). \label{eq:AFMagnon}
\end{align}
For the spin-nematic fluctuations we find
\begin{align}
    \frac{\delta \mathcal{Q}_{xx}^{(1)}}{\sqrt{\bar n}} &= \sqrt{1+f_z} \Re \left( \mathrm{e}^{-\mathrm{i} \chi_+} (\phi_- - \tfrac{1}{3} \phi_+) \right) \\
    &+\sqrt{1-f_z} \Re \left(  \mathrm{e}^{-\mathrm{i} \chi_-} (\phi_+ - \tfrac{1}{3} \phi_-) \right), \\
    \frac{\delta \mathcal{Q}_{yy}^{(1)}}{\sqrt{\bar n}}  &= \sqrt{1+f_z} \Re \left( \mathrm{e}^{-\mathrm{i} \chi_+} (\phi_- + \tfrac{1}{3} \phi_+) \right) \\
    &- \sqrt{1-f_z} \Re \left(  \mathrm{e}^{-\mathrm{i} \chi_-} (\phi_+ + \tfrac{1}{3} \phi_-) \right), \\
    \frac{\delta \mathcal{Q}_{zz}^{(1)}}{\sqrt{\bar n}} &= \frac{2}{3} \frac{\delta n^{(1)}}{\sqrt{\bar n}}, \\
    \frac{\delta \mathcal{Q}_{xy}^{(1)}}{\sqrt{\bar n}}  &=  \sqrt{1+f_z} \Im (\mathrm{e}^{- \mathrm{i} \chi_+} \phi_-) - \sqrt{1-f_z} \Im (\mathrm{e}^{- \mathrm{i} \chi_-}  \phi_+) , \\
    \frac{\delta \mathcal{Q}_{xz}^{(1)}}{\sqrt{\bar n/2}}  &= \sqrt{1+f_z} \Re ( \mathrm{e}^{-\mathrm{i} \chi_+} \phi_0) - \sqrt{1-f_z} \Re ( \mathrm{e}^{-\mathrm{i} \chi_-} \phi_0), \\
    \frac{\delta \mathcal{Q}_{yz}^{(1)}}{\sqrt{\bar n/2}}  &= \sqrt{1+f_z} \Im ( \mathrm{e}^{-\mathrm{i} \chi_+} \phi_0) + \sqrt{1-f_z} \Im ( \mathrm{e}^{-\mathrm{i} \chi_-} \phi_0). 
\end{align}

\subsection{Quadratic fluctuations}
The background-independent quadratic fluctuations read explicitly
\begin{equation}
    \begin{aligned}
        \delta n^{(2)} &= \tfrac{1}{2} (\lvert \phi_+ \rvert^2 + \lvert \phi_0 \rvert^2 + \lvert \phi_- \rvert^2 ), \\[4pt]
        \delta F_x^{(2)} &= \tfrac{1}{\sqrt{2}}  \phi_0^\mathrm{re} (\phi_+^\mathrm{re} + \phi_-^\mathrm{re} )   +  \tfrac{1}{\sqrt{2}} \phi_0^\mathrm{im} (\phi_+^\mathrm{im} + \phi_-^\mathrm{im}),  \\[4pt]
        \delta F_y^{(2)} &= \tfrac{1}{\sqrt{2}} \phi_0^\mathrm{im} (\phi_+^\mathrm{re} - \phi_-^\mathrm{re}) + \tfrac{1}{\sqrt{2}} \phi_0^\mathrm{re}  (\phi_-^\mathrm{im} - \phi_+^\mathrm{im}) , \\[4pt]
        \delta F_z^{(2)} &= \tfrac{1}{2} (\lvert \phi_+ \rvert^2 - \lvert \phi_- \rvert^2), \\[2pt]
        \delta \mathcal{Q}_{xx}^{(2)} &= \tfrac{1}{6} (2 \lvert \phi_0 \rvert^2 - \lvert \phi_+ \rvert^2 - \lvert \phi_- \rvert^2) + (\phi_+^\mathrm{im} \phi_-^\mathrm{im} + \phi_+^\mathrm{re} \phi_-^\mathrm{re}), \\[4pt]
        \delta \mathcal{Q}_{yy}^{(2)} &= \tfrac{1}{6} (2 \lvert \phi_0 \rvert^2 - \lvert \phi_+ \rvert^2 - \lvert \phi_- \rvert^2) - (\phi_+^\mathrm{im} \phi_-^\mathrm{im} + \phi_+^\mathrm{re} \phi_-^\mathrm{re}), \\[4pt]
        \delta \mathcal{Q}_{zz}^{(2)} &= \tfrac{2}{3} \delta n^{(2)} - \lvert \phi_0 \rvert^2 , \\[4pt]
        \delta \mathcal{Q}_{xy}^{(2)} &= \phi_+^\mathrm{re} \phi_-^\mathrm{im} - \phi_+^\mathrm{im} \phi_-^\mathrm{re},  \\[4pt]
        \delta \mathcal{Q}_{xz}^{(2)} &= \tfrac{1}{\sqrt{2}}  \phi_0^\mathrm{re} (\phi_+^\mathrm{re} - \phi_-^\mathrm{re}) + \phi_0^\mathrm{im} (\phi_+^\mathrm{im} - \phi_-^\mathrm{im}), \\[4pt]
        \delta \mathcal{Q}_{yz}^{(2)} &= \tfrac{1}{\sqrt{2}}  \phi_0^\mathrm{im} (\phi_+^\mathrm{im} + \phi_-^\mathrm{im}) - \tfrac{1}{\sqrt{2}}  \phi_0^\mathrm{re} (\phi_+^\mathrm{re} + \phi_-^\mathrm{re}). 
    \end{aligned} 
    \label{eq:QuadraticFluctuations}
\end{equation}

\section{Excitations of explicit symmetry-breaking mean field states}
\label{sec:ExplicitSymmetryBreakingDetailed}

\subsection{Polar phase}
At finite magnetic field, $p \neq 0$ and $q \neq 0$, the potential terms evaluate to  
\begin{equation}
    \begin{aligned}
    &\delta U + V_\mathrm{ext} \delta n^{(2)} \\
    &= \bar n c_0 (\Re \bm{\varphi}_\parallel)^2 + \bar n c_1 (\Im \bm{\varphi}_\perp)^2 + \frac{1}{2} \langle \bm{\varphi}_\perp^*,\big( q - p \, \mathrm{i} \sigma_y \big) \bm{\varphi}_\perp \rangle.
    \end{aligned}
    \label{eq:PotentialTermsPolar}
\end{equation}
Hence, 
\begin{equation}
        \begin{aligned}
            \Gamma_2^\parallel &= \int_{t,\bm{r}} \bigg \lbrace \hbar \langle \Im \bm{\varphi}_\parallel , \partial_t \Re \bm{\varphi}_\parallel \rangle + \frac{\hbar^2}{4M} \langle \bm{\varphi}^*_\parallel, \bm{\nabla}^2 \bm{\varphi}_\parallel \rangle \\
            &\qquad - c_0 \bar n (\Re \bm{\varphi}_\parallel)^2  \rangle \bigg \rbrace,
        \end{aligned}
\end{equation}
whereas
\begin{equation}
    \begin{aligned}
        \Gamma_2^\perp &= \int_{t,\bm{r}} \bigg \lbrace \hbar \langle \Im \bm{\varphi}_\perp, \partial_t \Re \bm{\varphi}_\perp \rangle + \frac{\hbar^2}{4M} \langle \bm{\varphi}_\perp^*, \bm{\nabla}^2 \bm{\varphi}_\perp \rangle \\
        &\qquad - c_1\bar n (\Im \bm{\varphi}_\perp)^2 + \frac{p}{2} \, \langle \bm{\varphi}_\perp^*, (\mathrm{i} \sigma_y) \, \bm{\varphi}_\perp \rangle - \frac{q}{2} \langle \bm{\varphi}_\perp^*, \bm{\varphi}_\perp \rangle \bigg \rbrace.
    \end{aligned}
    \label{eq:SpinorFluctuationPolar}
\end{equation}
At thix point it is instructive to derive the dispersion relations of the phonon and the magnon.
To that end we introduce the inverse propagators $G_{\parallel, \perp}^{-1}(\omega,\mathrm{k})$ by expressing the effective action 
in the form 
\begin{equation}
    \Gamma_2^{\parallel} = - \frac{1}{2} \int_{\omega,\bm{k}} \langle \bm{\varphi}_{\parallel}^*, G_{\parallel}^{-1}(\omega,\mathrm{k}) \bm{\varphi}_{\parallel} \rangle
\end{equation}
(and similar for $\Gamma_2^{\perp}$).
One finds the inverse phonon propagator
\begin{equation}
    G_\parallel^{-1}(\omega,\mathrm{k}) = \mqty(E_k + 2 \bar n c_0 & - \mathrm{i} \hbar \omega \\ \mathrm{i} \hbar \omega & E_k ),
\end{equation}
and the inverse magnon propagator
\begin{equation}
    G_0^{-1}(\omega,\mathrm{k}) = \mqty(E_k& - \mathrm{i} \hbar \omega \\ \mathrm{i} \hbar \omega & E_k + 2 \bar n c_1 + q),
\end{equation}
where we introduced the atomic kinetic energy $E_k = \hbar^2 k^2 /2M$. 
From their poles we find the dispersion relations
\begin{align}
    \hbar^2 \omega_\parallel^2(k) &= E_k (E_k + 2 \bar n c_0), \\
    \hbar^2 \omega_\perp^2(k) &= E_k (E_k + 2\bar n c_1 + q),
\end{align}
where we introduce the atomic kinetic energy $E_k = \hbar^2 k^2 /2M$.
Hence, both the phonon and the magnon carry the Bogoliubov dispersion relation, as it is well-known \cite{Kawaguchi2012}.

To derive the emergent spacetime in the magnonic sector one needs to integrate out $\Im \bm{\varphi}_\perp$ via evaluation of \eqref{eq:SpinorFluctuationPolar} on the field equations
\begin{align}
    \hbar \partial_t \Re \bm{\varphi}_\perp &= \left(2 \bar n c_1 + q - p \, \epsilon - \frac{\hbar}{2M} \bm{\nabla}^2 \right) \Im \bm{\varphi}_\perp, \\
    - \hbar \partial_t \Im \bm{\varphi}_\perp &= \left(q - p \, \epsilon -\frac{\hbar}{2M} \bm{\nabla}^2 \right) \Re \bm{\varphi}_\perp,
\end{align}
in the regime where the dispersion relation is linear, $\hbar^2 \bm{\nabla}^2/ 2M \ll (2 \bar n c_1 + q)$.
One finds 
\begin{align}
\begin{split}
        \Gamma_2^\perp &= - \frac{\hbar^2}{4M} \int_{t,\bm{r}} \bigg \lbrace - \frac{2M}{2 \bar n c_1 + q} (\partial_t \Re \bm{\varphi}_\perp)^2 \\
        &\qquad \qquad  + (\bm{\nabla} \Re \bm{\varphi}_\perp)^2 +  \frac{2M}{\hbar^2} q \, (\Re \bm{\varphi}_\perp)^2  \bigg \rbrace.
\end{split}
\end{align}

\subsection{Ferromagnetic phase}
\label{sec:ferro}
For the positively polarized case, $\bar f_z = 1$, the effective potential for the fluctuations reads
\begin{equation}
    \begin{aligned}
        &U(\bar \Phi + \delta \Phi) - U(\bar \Phi) + V_\mathrm{ext} \delta n^{(2)} \\
        &= \bar n\left( c_0 + c_1 \right) \left( \phi_+^\text{re} \right)^2 
        + \frac{1}{2} \left( p -q \right) \left( \left(\phi_0^\text{re} \right)^2 
        + \left( \phi_0^\text{im} \right)^2 \right) \\
        &\qquad + \left( p - \bar n c_1 \right) \left( \left( \phi_-^\text{re} \right)^2 + \left( \phi_-^\text{im} \right)^2 \right),
    \end{aligned}
\end{equation}
in the ferromagnetic phase. Here, the quadratic action for the fluctuating fields can be written as
\begin{align}
    \begin{split}
        \Gamma_2 &=     \int_{t,\bm{r}} \bigg \lbrace 
        - \hbar \phi_+^\text{re} \partial_t \phi_+^\text{im} 
        - \hbar \phi_0^\text{re} \partial_t \phi_0^\text{im} 
        - \hbar \phi_-^\text{re} \partial_t \phi_-^\text{im} \\
        &-  \frac{\hbar^2}{4M} \left( (\bm{\nabla} \phi_-^\mathrm{im})^2 + (\bm{\nabla} \phi_+^\mathrm{im})^2\right) - \phi_-^\text{im} \left( p - \bar n c_1 \right) \phi_-^\text{im}  \\
        &- \frac{1}{2} \phi_0^\text{im} \left( p - q - \frac{\hbar^2}{2M} \bm{\nabla}^2  \right) \phi_0^\text{im} \\
        & - \frac{1}{2} \phi_0^\text{re} \left( p - q - \frac{\hbar^2}{2M} \bm{\nabla}^2  \right) \phi_0^\text{re} \\
        & -  \phi_-^\text{re} \left(  p - \bar n c_1 - \frac{\hbar^2}{4M} \bm{\nabla}^2  \right) \phi_-^\text{re} \\
        & -  \phi_+^\text{re} \left(  \bar n \left( c_0 + c_1 \right) - \frac{\hbar^2}{4M} \bm{\nabla}^2 \right) \phi_+^\text{re}
        \bigg \rbrace.
    \end{split}
    \label{eq:FluctuationsActionFerro}
\end{align}
Let us also here derive the dispersion relations of the modes.
To that end we again introduce the inverse propagator $G^{-1}(\omega,\mathrm{k})$ by expressing the effective action 
in the form 
\begin{equation}
    \Gamma_2 = - \frac{1}{2} \int_{\omega,\bm{k}} (\phi^\mathrm{re}_\mathfrak{m}, \phi^\mathrm{im}_\mathfrak{m})  G_{\mathfrak{m}\mathfrak{m}'}^{-1}(\omega,\mathrm{k}) \phi_\mathfrak{m'}  (\phi^\mathrm{re}_\mathfrak{m}, \phi^\mathrm{im}_\mathfrak{m})^\mathrm{T}.
\end{equation}
Since the modes do not interact we have the block-diagonal form 
\begin{equation}
    G^{-1}(\omega,\mathrm{k}) = \mathrm{diag}(G_+^{-1}(\omega,\mathrm{k}), G_0^{-1}(\omega,\mathrm{k}), G_-^{-1}(\omega,\mathrm{k}))
\end{equation}
with the density block
\begin{equation}
    G_+^{-1}(\omega,\mathrm{k}) = \mqty(E_k + \bar n (c_0 + c_1) & - \mathrm{i} \hbar \omega \\ \mathrm{i} \hbar \omega & E_k ),
\end{equation}
the spin block 
\begin{equation}
    G_0^{-1}(\omega,\mathrm{k}) = \mqty(E_k + p-q & - \mathrm{i} \hbar \omega \\ \mathrm{i} \hbar \omega & E_k + p-q ),
\end{equation}
and the nematic block 
\begin{equation}
        G_-^{-1}(\omega,\mathrm{k}) = \mqty(E_k + 2(p - \bar n c_1) & - \mathrm{i} \hbar \omega \\ \mathrm{i} \hbar \omega & E_k + 2(p - \bar n c_1) ).
\end{equation} 
From the poles of the propagator, $\det G_\mathfrak{m}^{-1}(\omega_\mathfrak{m},\mathrm{k}) = 0$, we can read off the dispersion relations 
\begin{align}
    \hbar^2 \omega_+^2(k) &= E_k [E_k + \bar n (c_0 + c_1)] \label{eq:OmegaPlusFerro} \\[2pt]
    \hbar^2 \omega_0^2(k) &= (E_k + p - q)^2 \label{eq:OmegaNullFerro}  \\[2pt]
    \hbar^2 \omega_-^2(k) &= (E_k + 2p - 2 \bar n c_1)^2  \label{eq:OmegaMinusFerro}  
\end{align}
which can also be found in \cite{Kawaguchi2012}.
The density modes ($\phi_+$) follow a Bogoliubov dispersion relations whereas the spin ($\phi_0$) and nematic ($\phi_-$) modes have gapped, quadratic dispersion relations.
Let us proceed by deriving the field equations.
Functional differentiation of \cref{eq:FluctuationsActionFerro} yields
\begin{align}
    \hbar \partial_t \phi_0^\mathrm{re} &=  (p-q) \phi_0^\mathrm{im} - \frac{\hbar^2}{2M} \bm{\nabla}^2 \phi_0^\mathrm{im}, \\
    -\hbar \partial_t \phi_0^\mathrm{im} &=  (p-q) \phi_0^\mathrm{re} - \frac{\hbar^2}{2M} \bm{\nabla}^2 \phi_0^\mathrm{re},
\end{align}
and
\begin{align}
\hbar \partial_t \phi_-^\mathrm{re}  &=  2 (p - \bar n c_1) \phi_-^\mathrm{im} - \frac{\hbar^2}{2M} \bm{\nabla}^2 \phi_-^\mathrm{im}, \label{eq:PhiMRe_EoMFerro} \\
-\hbar \partial_t \phi_-^\mathrm{im} &=  2 (p -  \bar n c_1) \phi_-^\mathrm{re} - \frac{\hbar^2}{2M} \bm{\nabla}^2 \phi_-^\mathrm{re},
\label{eq:PhiMIm_EoMFerro}
\end{align}
as well as
\begin{align}
\hbar \partial_t \phi_+^\mathrm{re}  &=  - \frac{\hbar^2}{2M} \bm{\nabla}^2 \phi_+^\mathrm{im},
 \label{eq:PhiPRe_EoMFerro} \\
 - \hbar \partial_t \phi_+^\mathrm{im} &= 2 \bar n (c_0 + c_1) \phi_+^\mathrm{re} - \frac{\hbar^2}{2M} \bm{\nabla}^2 \phi_+^\mathrm{re}. 
\label{eq:PhiPIm_EoMFerro}
\end{align}
To arrive at an effective FLRW-spacetime, it is now crucial to integrate out $\phi_0^\mathrm{re}$ and $\phi_\pm^\mathrm{re}$ in the low-momentum regimes $E_k \ll p-q$ for the spin modes ($\phi_0$) and $E_k \ll 2(p - \bar n c_1)$ for the nematic modes ($\phi_-$).
This leads to
\begin{equation}
    \begin{aligned}
            &\Gamma_2 = - \frac{\hbar^2}{2} \int_{t,\bm{r}} \bigg \lbrace - \frac{\left( \partial_t \phi_+^\mathrm{im} \right)^2}{2 \bar n \left( c_0 + c_1 \right)}  + \frac{\left( \bm{\nabla} \phi_+^\mathrm{im} \right)^2}{2M}  \\[2pt]
            &- \frac{\left( \partial_t \phi_0^\mathrm{im} \right)^2}{p - q}  + \frac{\left( \bm{\nabla} \phi_0^\mathrm{im} \right)^2 }{2M} 
            + \frac{\left( p - q \right)}{\hbar^2}  \left( \phi_0^\mathrm{im} \right)^2 \\[2pt]
            &- \frac{\left( \partial_t \phi_-^\mathrm{im} \right)^2 }{2(p - \bar n c_1)} + \frac{\left( \bm{\nabla} \phi_-^\mathrm{im} \right)^2}{2M} 
            + \frac{2\left( p - \bar n c_1 \right)}{\hbar^2} \left( \phi_-^\mathrm{im} \right)^2 \bigg \rbrace. \\[6pt]
            &\equiv \Gamma_2^+ + \Gamma_2^0 + \Gamma_2^-,
    \end{aligned}
    \label{eq:acoustic-ferro}
\end{equation}

\subsection{Anti-Ferromagnetic phase}
\label{ap:anti}
In the anti-ferromagnetic phase, one finds
    \begin{equation}
        \begin{aligned}
            & U(\bar \Phi + \delta \Phi) - U(\bar \Phi) + V_\mathrm{ext} \delta n^{(2)} \\[4pt]
            &=  \frac{1}{2} (q_- - q) (\phi_0^\text{re})^2 + \frac{1}{2} (q_+ - q) (\phi_0^\text{im})^2 \\[4pt]
            &+ \frac{\bar n c_0}{2}\big( \sqrt{1+ f_z} \, \mathrm{Re}[\mathrm{e}^{- \mathrm{i} \chi_+} \phi_+] + \sqrt{1 - f_z} \, \mathrm{Re}[\mathrm{e}^{- \mathrm{i} \chi_-} \phi_-] \big)^2 \\[4pt]
            &+ \frac{\bar n c_1}{2}\big( \sqrt{1+ f_z} \, \mathrm{Re}[\mathrm{e}^{- \mathrm{i} \chi_+} \phi_+] - \sqrt{1 - f_z} \, \mathrm{Re}[\mathrm{e}^{- \mathrm{i} \chi_-} \phi_-] \big)^2 \\[4pt]
        \end{aligned}
    \end{equation}
where we replaced the external potential by \cref{eq:eom-anti}.
The nematic sector contains the field $\phi_0$ which is described by the effective action 
\begin{align}
    \begin{split}
        &\Gamma_2^0 = \int \diff t\diff^2 \bm{r} \lbrace - \hbar \phi_0^\text{re}\partial_t \phi_0^\text{im} + \frac{\hbar^2}{4M} \phi_0^\mathrm{im} \bm{\nabla}^2 \phi_0^\mathrm{im} \\[3pt] 
       &- \frac{\hbar^2}{4M} (\bm{\nabla} \phi_0^\mathrm{re})^2 - \frac{1}{2} \phi_0^\mathrm{re} (q_- - q) \phi_0^\mathrm{re} - \frac{1}{2} \phi_0^\mathrm{im} (q_+ - q) \phi_0^\mathrm{im}  \rbrace,
    \end{split}
    \label{eq:FluctuationsActionAnti}
\end{align}
which generates the field equations
\begin{align}
    \hbar \partial_t \phi_0^\mathrm{re} &= \left( q_+ - q \right) \phi_0^\mathrm{im} - \frac{\hbar^2}{2M} \bm{\nabla}^2 \phi_0^\mathrm{im},  \label{eq:Phi0Re_EoMAF} \\
    -\hbar \partial_t \phi_0^\mathrm{im}&= \left( q_- - q \right) \phi_0^\mathrm{re} - \frac{\hbar^2}{2M} \bm{\nabla}^2 \phi_0^\mathrm{re}, \label{eq:Phi0Im_EoMAF}
\end{align}
and can be described by the inverse propagator 
\begin{equation}
    G_0^{-1}(\omega,\mathrm{k}) = \mqty(E_k + q_- -q & - \mathrm{i} \hbar \omega \\ \mathrm{i} \hbar \omega & E_k + q_+ - q ),
\end{equation}
which has a pole at the dispersion relation 
\begin{equation}
    \hbar^2 \omega_0^2 = (E_k + q_- - q) (E_k + q_+ - q).
\end{equation}
Integrating out $\phi_0^\mathrm{re}$ in the low-momentum regime $E_k \ll q_- - q$, we find
\begin{equation}
    \begin{aligned}
        \Gamma_2^0 &= -\frac{\hbar^2}{2} \int_{t,\bm{r}} \bigg \lbrace - \frac{(\partial_t \phi_0^\mathrm{im})^2}{q_- - q}   + \frac{ (\bm{\nabla} \phi_0^\mathrm{im})^2}{2M} \\ &+ \frac{q_+ - q}{\hbar^2} (\phi_0^\mathrm{im})^2 \bigg \rbrace.    \end{aligned}
\end{equation}
For the phonon-magnon sector, it is convenient to introduce 
\begin{equation}
    \varphi_\pm = \frac{1}{\sqrt{2}} \left(\mathrm{e}^{- \mathrm{i} \chi_+} \phi_+ \pm \mathrm{e}^{- \mathrm{i} \chi_-} \phi_-\right),
    \label{eq:PhononMagnonFieldsFree}
\end{equation}
which represents phonons ($+$) and magnons ($-$) if $f_z = 0$ (cref.\@ \cref{eq:AFPhonon,eq:AFMagnon}).
The case of $f_z \neq 0$ will be treated as a perturbation in the following.
For the effective action in the phonon-magnon-sector we find
\begin{equation}
    \begin{aligned}
        \Gamma_2^\mathrm{\pm} &=  \int_{t,\bm{r}} \big \lbrace  \varphi_-^\mathrm{im} \hbar \partial_t \varphi_-^\mathrm{re} + \varphi_+^\mathrm{im} \hbar \partial_t \varphi_+^\mathrm{re} \\[2pt]
        &- \frac{\hbar^2}{4M} \left( (\bm{\nabla} \varphi_-^\mathrm{im})^2 + (\bm{\nabla} \varphi_-^\mathrm{re})^2 + (\bm{\nabla} \varphi_+^\mathrm{im})^2 + (\bm{\nabla} \varphi_+^\mathrm{re})^2  \right) \\[2pt]
        &- \bar n c_0 (\varphi_+^\mathrm{re})^2 - \bar n c_1 (\varphi_-^\mathrm{re})^2 + f_z \bar n (c_0 + c_1)  \varphi_+^\mathrm{re} \, \varphi_-^\mathrm{re}  \\[2pt]
        &+ \frac{1}{2} \bar n (c_0 - c_1) (\sqrt{1-f_z^2}-1) \left((\varphi_+^\mathrm{re})^2 - (\varphi_-^\mathrm{re})^2 \right) \big \rbrace,
    \end{aligned}
\end{equation}
which generates the field equations
\begin{align}
    \hbar \partial_t \varphi_\pm^\mathrm{re} &= - \frac{\hbar^2 \bm{\nabla}^2}{2M} \varphi_\pm^\mathrm{im}, \label{eq:PhiPMRe_EoMAF} \\
    -\hbar \partial_t \varphi_\pm^\mathrm{im}  &= 2 \bar n c_{0,1} \varphi_\pm^\mathrm{re} \pm 2 \bar n (c_0 - c_1) (1 - \sqrt{1-f_z^2}) \varphi_\pm^\mathrm{re}   \nonumber \\
    &+ f_z \bar n (c_0 + c_1) \varphi_\mp^\mathrm{re} - \frac{\hbar^2 \bm{\nabla}^2}{2M} \varphi_\pm^\mathrm{re}, \label{eq:PhiPMIm_EoMAF}
\end{align}
and is associated to the inverse propagator
\begin{equation}
    G^{-1}(\omega,\mathrm{k}) = \mqty(G_{\mathrm{DD}}^{-1} & G_{\mathrm{DS}}^{-1} \\G_{\mathrm{DS}}^{-1} & G_{\mathrm{SS}}^{-1} )
\end{equation}
with the Bogoliubov blocks
\begin{equation}
    G_{\mathrm{DD},\mathrm{SS}}(\omega,\mathrm{k}) = \mqty(
    E_k + \mu_\mathrm{D,S}^\mathrm{eff} & - \mathrm{i} \hbar \omega \\
    \mathrm{i} \hbar \omega & E_k
    ),
\end{equation}
where we introduced effective chemical potentials
\begin{align}
    \mu_\mathrm{D}^\mathrm{eff} = 2 \bar n c_0 + \bar n (c_0 - c_1) (1 - \sqrt{1-f_z^2}),  \\
    \mu_\mathrm{S}^\mathrm{eff} = 2 \bar n c_1 - \bar n (c_0 - c_1) (1 - \sqrt{1-f_z^2}). 
\end{align}
In absence of an external magnetic field ($f_z = 0$) we have $\mu_\mathrm{D,S}^\mathrm{eff} = 2 \bar n c_{0,1}$ and the eigenmodes of the sytem are given by \cref{eq:PhononMagnonFieldsFree}.
The effect of an external magnetic field ($f_z \neq 0$) on the phonons and magnons is captured by the effective chemical potentials.
However, one also has the mixing terms
\begin{equation}
    G_{\mathrm{DS}}^{-1}(\omega,\mathrm{k}) 
    = \mqty(
    f_z \bar n (c_0 + c_1)   & 0 \\
    0 & 0),
\end{equation}
such that the exact eigenmodes are given by a superposition of the phonon and magnon modes \cite{Kawaguchi2012}.
More concretely, the exact eigenmodes have the dispersion relation
\begin{equation}
    \begin{aligned}
        \omega_\pm^2(k) &= \frac{\omega_\mathrm{D}^2(k) + \omega_\mathrm{S}^2(k)}{2} \\
        &\pm \sqrt{ \frac{1}{4}\left(\omega_\mathrm{D}^2(k) - \omega_\mathrm{S}^2(k)\right)^2 + \frac{f_z^2}{\hbar^2}  E_k^2 \big(\bar n (c_0 + c_1)\big)^2 },
    \end{aligned}
\end{equation}
which is the avoided crossing of the perturbed phonon and magnon branches,
\begin{equation}
    \hbar^2 \omega_\mathrm{D,S}^2(k) = E_k (E_k + \mu_\mathrm{D,S}).
\end{equation}
Equivalently, we can write
\begin{equation}
    \begin{aligned}
        \omega_\pm^2(k) &= \frac{\omega_{\mathrm{D},0}^2(k) + \omega_{\mathrm{S},0}^2(k)}{2} \\
        &\pm \sqrt{ \frac{1}{4}\left( \omega_{\mathrm{D},0}^2(k) - \omega_{\mathrm{S},0}^2(k) \right)^2 + \frac{f_z^2}{\hbar^2} E_k^2 (2 \bar n c_0) (2 \bar n c_1)},
    \end{aligned}
\end{equation}
as an avoided crossing of the \emph{unpertubed} phonon and magnon branches, 
\begin{equation}
    \begin{aligned}
        \hbar^2 \omega_\mathrm{D,0}^2(k) &= E_k (E_k + 2 \bar n c_0), \\
        \hbar^2 \omega_\mathrm{S,0}^2(k) &= E_k (E_k + 2 \bar n c_1).
    \end{aligned}
\end{equation}

\section{Proca fields minimally coupled to homegeneous and isotropic spacetimes}
\label{sec:ProcaField}

We consider a Proca field which is minimally coupled to a curved spacetime in terms of the effective action $\Gamma = \int \mathrm{d}^{D+1} x \sqrt{g} \mathcal{L}$, with the Lagrangian density \cite{Graham2016,Heisenberg2019}
\begin{equation}
        \mathcal{L} = - \frac{1}{4} F_{\mu \nu} F^{\mu \nu} - \frac{1}{2} m_A^2 g^{\mu \nu}  A_\mu A_\nu,
\end{equation}
where we introduced the field strength tensor 
\begin{equation}
    F_{\mu \nu} = \nabla_\mu A_\nu - \nabla_\nu A_\mu = \partial_\mu A_\nu - \partial_\nu A_\mu,
\end{equation}
and $\nabla_\mu$ as the covariant derivative. 
The equations of motion take the form 
\begin{equation}
    \nabla_\mu F^{\mu \nu} = m_W^2 A^\nu
\end{equation}
where we used that 
\begin{equation}
    \nabla_\mu F^{\mu \nu} = \frac{1}{\sqrt{g}} \partial_\mu \left( \sqrt{g} F^{\mu \nu} \right)
\end{equation}
since $F_{\mu \nu}$ is antisymmetric.
More explicitly, we have
\begin{equation}
    \nabla_\mu F^{\mu \nu} = \frac{1}{\sqrt{g}} \partial_\mu \left(\sqrt{g} \, \partial^\mu A^\nu \right) - \partial^\nu \partial_\mu A^\mu - \frac{\partial_\mu \sqrt{g}}{\sqrt{g}}  \partial^\nu A^\mu.
\end{equation}
In a gauge where $A_0 = 0$ and $\partial_i A^i = 0$ is realized on the solution to the equations of motion, one has
\begin{equation}
    \nabla_\mu F^{\mu i} = \frac{1}{\sqrt{g}} \partial_\mu \left(\sqrt{g} \, g^{\mu \nu} \partial_\nu A^i \right) - \frac{\partial_j \sqrt{g}}{\sqrt{g}}  g^{ik} \partial_k A^j.
\end{equation}
For homegeneous and isotropic spacetimes, one has $\partial_j \sqrt{g} = 0$, such that the field equation becomes
\begin{equation}
    \nabla_\mu F^{\mu i} = \frac{1}{\sqrt{g}} \partial_\mu \left(\sqrt{g} \, g^{\mu \nu} \partial_\nu A^i \right) = m_A^2 A^i.
\end{equation}

\bibliography{main}
\end{document}